\documentclass[12pt]{article}
\pdfoutput=1
\usepackage{jheppub}
\usepackage{amsmath}
\usepackage{MnSymbol}

\def\be{\begin{equation}}
\def\ee{\end{equation}}
\def\ba{\begin{aligned}}
\def\ea{\end{aligned}}
\def\ben{\begin{eqnarray}\displaystyle}
\def\een{\end{eqnarray}}

\def\nn{\nonumber}
\def\rd{{\rm d}}

\def\rd{{\rm d}}

\newcommand{\bal}{\begin{equation}\begin{aligned}}
\newcommand{\eal}{\end{aligned}\end{equation}}

\newcommand{\alg}[1]{\mathfrak{#1}}

\begin{document}
\begin{flushright} {DMUS--MP--13/19, PUPT-2456, LPTENS 13/23}\end{flushright}

\begin{center}
\vspace*{5mm}
{\Large\sf
{5D partition functions, $q$-Virasoro systems and integrable spin-chains}
}

\vspace*{5mm}
{\large Fabrizio Nieri${}^\clubsuit$, Sara Pasquetti${}^\clubsuit$, Filippo Passerini${}^\diamondsuit$,\\ Alessandro Torrielli${}^\clubsuit$}

\vspace*{5mm}
       ${}^\clubsuit$ Department of Mathematics, University of Surrey, Guildford, Surrey, GU2 7XH, UK
       
       ${}^\diamondsuit$ Department of Physics, Princeton University, Princeton, NJ 08544

and

Laboratoire de Physique Th\'{e}orique\footnote{Unit\'{e} Mixte du CNRS et de l'Ecole Normale Sup\'{e}rieure associ\'{e}e \`{a} l'Universit\'{e} Pierre et Marie Curie 6, UMR 8549.},  Ecole Normale Sup\'{e}rieure, 75005 Paris, France
\vspace*{5mm}

{\tt fb.nieri@gmail.com, sara.pasquetti@gmail.com, filipasse@gmail.com, a.torrielli@surrey.ac.uk}

\vspace*{9mm}
{\bf Abstract} 
\end{center}
\vspace*{1mm}
\noindent We analyze $\mathcal{N} = 1$ theories on $S^5$ and $S^4 \times S^1$, showing how their partition functions can be written in terms of a set of fundamental 5d holomorphic blocks. We demonstrate
that, when the 5d mass parameters are analytically continued to suitable values, the $S^5$ and
$S^4 \times S^1$ partition functions degenerate to those for $S^3$ and $S^2 \times S^1$. We explain
this mechanism via the recently proposed correspondence between 5d partition
functions and correlators with underlying $q$-Virasoro symmetry. From the  $q$-Virasoro 3-point
functions, we  axiomatically derive a set of associated reflection coefficients,
and show that they can be geometrically interpreted   in terms of Harish-Chandra $c$-functions
for quantum symmetric spaces. We  link these particular $c$-functions to the types appearing in the Jost functions
encoding the asymptotics of the scattering in integrable spin-chains, obtained taking different limits of the XYZ model to XXZ-type.
\newpage
\setcounter{tocdepth}{1}
\tableofcontents


\section{Introduction}
The study of supersymmetric gauge theories on compact manifolds has attracted much attention in recent years.
After the seminal work by Pestun \cite{Pestun:2007rz}, the method of supersymmetric localization has been applied to compute  partition functions of theories formulated on compact manifolds of various dimensions. A comprehensive approach to  rigid supersymmetry  in curved backgrounds has been also proposed   \cite{Festuccia:2011ws}.

In this paper we focus on 5d   $\mathcal{N}=1$ theories. 
Exact results for partition functions of $\mathcal{N}=1$ theories on $S^5$ and $S^4\times S^1$ were derived in \cite{Kallen:2012cs, Hosomichi:2012ek, Kallen:2012va, Kim:2012ava, Imamura:2012bm, Lockhart:2012vp, Kim:2012qf, Minahan:2013jwa} and \cite{Kim:2012gu,Terashima:2012ra,Iqbal:2012xm}. In the case of the squashed $S^5$, the partition function $Z_{S^5}$ was shown to localize to a matrix integral of classical,  1-loop and instanton contributions. The latter  in turn comprises of three copies of the equivariant instanton partition function on $\mathbb{R}^4\times S^1$  \cite{Nekrasov:2002qd,Nekrasov:2003rj} with an appropriate identification of the equivariant parameters, each copy  corresponding to the contribution at a fixed point of the Hopf fibration of $S^5$ over $\mathbb{C P}^2$.
The $S^4\times S^1$ case is similar with the instanton partition function  consisting of the product of two copies of the equivariant instanton partition function on $\mathbb{R}^4\times S^1$, corresponding to the fixed points at the north and south poles of $S^4.$ 

Our first  result is the observation that, by manipulating the classical and 1-loop part to a form which respects the symmetry dictated by the gluing of the instanton factors,  it is possible to rewrite  $Z_{S^5}$ and $Z_{S^4\times S^1}$ in terms of the same fundamental building blocks, which we name 5d holomorphic blocks ${\cal  B}^{5d} $.
 In formulas:
\be
\nn
Z_{S^5}=\int d \sigma ~\Big|\Big| {\cal  B}^{5d} \Big|\Big|^3_{S}\,,\qquad Z_{S^4\times S^1}=\int d \sigma ~\Big|\Big| {\cal  B}^{5d} \Big|\Big|^2_{id}\,,
\ee
where the brackets $\Big|\Big| \ldots \Big|\Big|^3_{S}$ and $\Big|\Big| \ldots \Big|\Big|^2_{id}$  glue respectively three and two 5d holomorphic blocks, as described in details in the main text.   This result  is very reminiscent of the 3d case   where $S^2\times S^1$ and $S^3$ partition functions were shown to  factorize in terms of the same set of building blocks, named 3d holomorphic blocks ${\cal  B}^{3d}_\alpha$, glued with different pairings \cite{Pasquetti:2011fj,hb}:\footnote{For a  proof  of the factorization property of 3-sphere partition functions see \cite{Alday:2013lba}.}
\be
\nn
Z_{S^3}=\sum_{\alpha}~\Big|\Big| {\cal  B}^{3d}_\alpha \Big|\Big|^2_{S}\,,\qquad Z_{S^2\times S^1}=
\sum_{\alpha}~\Big|\Big| {\cal  B}^{3d}_\alpha \Big|\Big|^2_{id}\,.
\ee
Holomorphic blocks in three dimensions were identified  with  solid tori or Melving cigars $M_q=D^2\times S^1$ partition functions, the subscripts  $id$, $S$ refer to the way blocks are fused  which was shown to be  consistent  with the decomposition of   $S^2\times S^1$ and $S^3$ in solid tori  glued by  $id$ and $S$ elements in $SL(2,\mathbb{Z})$. 
The index $\alpha$  labels the SUSY vacua of the semiclassical  $\mathbb{R}^2\times S^1$ theory but it also turns out to run over a basis of solutions to certain difference 
operators which annihilates 3d blocks and 3d partition functions \cite{Dimofte:2011py,Dimofte:2011ju,hb}.

In fact, the similarity between the structure of  5d and 3d partition functions is not just a coincidence, but it is due to a deep relation between the two theories.
For example we consider the 5d  $\mathcal{N}=1$ SCQCD with $SU(2)$ gauge group and four flavors  on $S^4\times S^1$ and $S^5$   and show that, when the masses are analytically continued to certain values,  5d partition functions  degenerate to 3d partition functions of the SQED with $U(1)$
gauge group and four flavors respectively on $S^2\times S^1$ and $S^3$. Schematically:
\be
\nn
Z_{S^5}^{SCQCD}=\int d \sigma ~\Big|\Big| {\cal  B}^{5d} \Big|\Big|^3_{S}\,\longrightarrow\, Z_{S^3}^{SQED}=\sum_{\alpha}~\Big|\Big| {\cal  B}^{3d}_\alpha \Big|\Big|^2_{S}\,,
\ee
and
\be
\nn
 Z_{S^4\times S^1}^{SCQCD}=\int d \sigma ~\Big|\Big| {\cal  B}^{5d} \Big|\Big|^2_{id} \,\longrightarrow\, Z_{S^2\times S^1}^{SQED}=
\sum_{\alpha}~\Big|\Big| {\cal  B}^{3d}_\alpha \Big|\Big|^2_{id}\,,
\ee
where the 5d $id$ and $S$ gluings reduce to the corresponding pairing for 3d theories introduced in \cite{hb}. The mechanisms that leads to this degeneration  is the fact  that, upon analytic continuation of the masses,
1-loop terms develop poles which  pinch the integration contour. Partition functions then receive contribution from the residues at the poles trapped along the integration contour.
This result is roughly the compact version of the degeneration of the instanton partition function to the vortex partition function of simple surface operators \cite{Alday:2009fs, Dimofte:2010tz,Kozcaz:2010af,Bonelli:2011fq}, since $S^2\times S^1$ and $S^3$ are codimension-2 defects inside respectively $S^4\times S^1$ and $S^5$.

The fact that when analytically continuing  flavor parameters partition functions degenerate
to sum of terms annihilated by difference operators is quite general, for example this is the case also for  the 4d superconformal index  \cite{Gaiotto:2012xa}. 
In the case of the degeneration of 4d instanton partition functions to surface operator vortex counting,  this mechanism  has been related, via the AGT correspondence, to  the analytic continuation of a primary operator to a degenerate primary in  Liouville correlators \cite{Alday:2009fs}. 
We can offer a similar interpretation of the rich structure of the  degeneration of 5d partition functions in the context  of the correspondence, proposed in  \cite{Nieri:2013yra},  between 5d partition functions and correlators with underlying symmetry given by a deformation of the  Virasoro algebra known as ${\cal V}ir_{q,t}$.
In  \cite{Nieri:2013yra} two families of such correlators, dubbed respectively $S$ and $id$-correlators, were constructed by means of the  bootstrap approach  using the explicit form of degenerate representations of ${\cal V}ir_{q,t}$ as well as an ansatz for  the pairings of the ${\cal V}ir_{q,t}$ chiral blocks, generalizing the familiar  holomorphic-antiholomorphic square. The two families of correlators differ indeed by the pairing of the blocks and are endowed respectively with   3-point functions $C_S$ and $C_{id}$.
In \cite{Nieri:2013yra} it was checked that  $Z_{S^5}$ and $Z_{S^4\times S^1}$ partition functions of   the 5d  $\mathcal{N}=1$ SCQCD with $SU(2)$ gauge group and $N_f=4$ are captured respectively  by 4-point $S$ and $id$-correlator on the 2-sphere.

In this paper we provide  another check showing that  1-point torus correlators capture
 the ${\cal N}=1^*$ $SU(2)$ theory, that is the theory of a vector multiplet coupled to one adjoint hyper.  We  also reinterpret and clarify the degeneration mechanisms of 5d partition functions
 in terms of fusion rules of degenerate ${\cal V}ir_{q,t}$ primaries. In particular, the pinching of the integration contour for the 5d partition functions is described in detail using the language of ${\cal V}ir_{q,t}$ correlators.

As pointed out  in { \cite{Zamolodchikov:1995aa}}, 3-point functions can also be used to define the {\it reflection coefficients}, which in turn encode very important information about the theory. In the case of  Liouville theory a semiclassical reflection coefficient,  which is perturbative in the Liouville coupling  constant $b_0$,
can be obtained from a ``first quantized"-type analysis. In the so-called {\it mini-superspace} approximation \cite{Seiberg}, one can solve the Schr\"odinger equation for the zero-mode of the Liouville field which scatters from an exponential barrier. The same equation appears when studying the radial part of the Laplacian in horospheric coordinates on the hyperboloid $SL(2,\mathbb{C})/SU(2)$ (Lobachevsky space) (see for instance \cite{Mironov,OlshanetskyRogov,OlshanetskyPerelomov} and further references in the main text).

In this case, the reflection coefficient can be expressed as a ratio of so-called Harish-Chandra $c$-functions for the root system of the underlying $\alg{sl}(2)$ Lie algebra. In fact, this procedure is purely group-theoretic 
 (see also \cite{GerasimovMarshakovOlshanetskyShatashvili,GerasimovMarshakovMorozovOlshanetskyShatashvili})  and can be generalized to the case of quantum group deformations. The results appear
to be given in terms of products of factors associated to the root decomposition of
the relevant (quantum) algebra. These factors, in turn, are
given in terms of certain special functions which are typically Gamma functions and
generalizations thereof  \cite{FreundZabrodinII}.
An important observation in \cite{GerasimovKharchevMarshakovMironovMorozovOlshanetsky} is that one can obtain the exact non-perturbative Liouville reflection coefficient  by considering the  affine  version of the
$\alg{sl}(2)$ algebra. While the non-affine version produces the semiclassical reflection
coefficient, the affinization  procedure, which includes the contributions from the affine root (equivalently, from the infinite  tower of affine levels), generates the exact non-pertubative result.
In a sense, the affinization procedure plays the role of a ``second quantization", restoring the non-perturbative dependence in the Liouville parameter $1/b_0$.

Inspired by this remarkable observation we checked whether our exact reflection coefficients, for the two types of geometries/correlator-pairings considered in \cite{Nieri:2013yra}, could be reproduced by a similar affinization procedure, starting from a suitable choice of Harish-Chandra $c$-functions. We found that this is indeed the case. Our reflection coefficients can be obtained by affinizing the ratio of Harish-Chandra functions given  in terms of 
$q$-deformed Gamma functions $\Gamma_q$ and  double Gamma functions $\Gamma_2$
for the $id$- and $S$-pairing, respectively. The $\Gamma_q$ function appears in the study of the quantum version of the Lobachevsky space, which involves the quantum algebra $U_q(\alg{sl}(2))$.\footnote{A lattice version of Liouville theory has been suggested in \cite{Bytsko}
to be related to the quantum algebra $U_q(\alg{sl}(2))$ via a certain {\it Baxterisation} procedure. This procedure can be thought of as a technique to consistently introduce the dependence on a spectral parameter in an otherwise constant R-matrix structure.} The appearance of the $\Gamma_2$ function is harder to directly relate to a specific quantum group construction, even thought it has been conjectured to be part of a hierarchy of integrable structures  \cite{FreundZabrodinII}.

It is quite remarkable to observe that the affinization mechanisms works for the situation in object in this paper precisely as it did for Liouville theory. In particular, it is highly non-trivial that such procedure is capable of restoring the $SL(3,\mathbb{Z})$ invariance of the 5d $S$-pairing  after the infinite resummation. We believe that this occurrence deserves further investigations.

To shed more light on the type of special
functions appearing in our reflection coefficients, we consider scattering of excitations in
integrable spin-chains, which are prototypical representatives in the universality classes
of 2-dimensional integrable systems. We  find that the special functions 
in our reflection coefficients are indeed typical of scattering processes between particle-like
excited states in different regimes of the XXZ system, obtained as limiting cases from the XYZ spin-chain. In particular, the {\it id}-pairing produces  $\Gamma_q$ functions that characterize the scattering in the {\it antiferromagnetic} regime of the XXZ spin-chain, while the {\it S}-pairing produces $\Gamma_2$ functions, related to the scattering on the XXZ chain in its {\it disordered} regime (and also, for instance, in the Sine-Gordon model). The link between reflection coefficients, Harish-Chandra functions and scattering matrices of integrable systems
is in essence the one anticipated in the work of Freund and Zabrodin
(see for instance \cite{FreundZabrodinII} and further references in the main text). In this sense, we can establish a direct connection between the axiomatic properties of the $q$-deformed systems underlying gauge theory partition functions, their semiclassical images consisting of quantum mechanical reflections from a potential barrier, their geometric pictures in terms of group theory data, and the exact analysis of 2-dimensional integrable hierarchies.

The underlying presence of the $q$-deformed Virasoro symmetry is strongly suggestive that the properties of the $q$-conformal blocks and of the emerging integrable systems should be intimately tied together. In particular, we expect that a significant  part will be played by the $q$-deformed Knizhnik-Zamolodchikov equation \cite{FrenkelReshetikhin,Lukyanov:1992sc}, and, correspondingly, by the integrable form-factor equations\footnote{We thank Samson Shatashvili for communication on this point.} \cite{Smirnov,LashkevichPugai}.

The appearance of the XXZ spin-chain in connection with reflection coefficients 
is not surprising since, as it was shown in   \cite{Nieri:2013yra}, the 3-point functions $C_S$ and $C_{id}$
were ultimately derived from 3d $\mathcal{N}=2$ theories whose SUSY vacua can be mapped to eigenstates of spin-chain Hamiltonians  \cite{ns1,ns2}. For further results in this direction see \cite{lc1,lc2,gk,gaddegukovputrovwalls}.

\section{5d holomorphic blocks}

In this section we study  5d ${\cal N}=1$ theories  formulated on  the  squashed   $S^5$ and  on $S^4\times S^1$.  We begin by recording the expressions obtained in the literature for the partition functions $Z_{S^5}$ and  $Z_{S^4\times S^1}$. We then show that $Z_{S^5}$ and $Z_{S^4\times S^1}$ can be decomposed in terms of    5d holomorphic blocks ${\cal B}^{5d}$. We also show how $S^5$ and   $S^4\times S^1$ partition functions degenerate to $S^3$ and   $S^2\times S^1$ partition functions when masses are analytically continued to suitable values.

\subsection{Squashed $S^5$ partition functions and 5d holomorphic blocks}

In a series of papers \cite{Kallen:2012cs, Hosomichi:2012ek, Kallen:2012va, Kim:2012ava, Imamura:2012bm, Lockhart:2012vp, Kim:2012qf, Minahan:2013jwa} the 5d ${\cal N}=1$ supersymmetric gauge theory has been formulated on 
 $S^5$ with squashing parameters $\omega_1,\omega_2,\omega_3$,
and the partition function has been shown to localize  to the integral over the zero-mode of the vector multiplet scalar $\sigma$ which takes value in  the Cartan subalgebra of the gauge group, {which we take to be $SU(N)$ with generators $T_a$ normalized as $\text{Tr}_R(T_a T_b)=C_2(R)\delta_{ab}$, with $C_2(F)=1/2$ for the fundamental.} The integrand includes a classical factor
${ Z}_{\text{cl}}(\sigma)$, a 1-loop factor ${ Z}_{\text{1-loop}}(\sigma,\vec M) $ and 
an instanton factor ${ Z}_{\text{inst}}(\sigma, \vec M)$
\be\label{s5}
Z_{S^5}=\int d \sigma ~{ Z}_{\text{cl}}(\sigma) { Z}_{\text{1-loop}}(\sigma,\vec M)  { Z}_{\text{inst}}(\sigma,\vec M) \,,
\ee
 where we schematically denote by $\vec M$ all  mass parameters of the theory and the explicit expressions of the factors depend on the field content  of the 5d ${\cal N}=1$ theory under consideration.\\

The instanton partition function ${ Z}_{\text{inst}}(\sigma,\vec M)$ receives  contributions 
from the three fixed points of the Hopf fibration over the $\mathbb{CP}^2$ base, and,
as show in  \cite{Lockhart:2012vp, Kim:2012qf},  takes the following factorized form: 
\be
 Z_\text{inst}(\sigma,\vec M) =\Big|\Big|{\cal Z}_\text{inst}\Big|\Big|^3_{S} \,,
\ee
where ${\cal Z}^{}_\text{inst}$ coincides with  the equivariant instanton partition function 
on ${\mathbb R}^4\times S^1$ \cite{Nekrasov:2002qd, Nekrasov:2003rj} with Coulomb and mass parameters appropriately rescaled and with equivariant parameters $\epsilon_1= \frac{e_1}{e_3}$
and $\epsilon_2= \frac{e_2}{e_3}$:
\be 
\label{cinst}
 {\cal Z}_\text{inst}=  {\cal Z}^{\mathbb{R}^4\times S^1}_\text {inst}\left(\frac{i \sigma}{e_3}, \frac{ \vec m}{e_3}; 
 \frac{e_1}{e_3},\frac{e_2}{e_3}\right)\,,
\ee
where  
\be
m_j=i M_j + E/2\,, \qquad j=1,\cdots, N_f\,,
\label{newmass}
\ee
and $E=\omega_1+\omega_2+\omega_3$.
We also  introduced the notation:
\ben\label{sl3}
\Big|\Big| (\dots)\Big|\Big|^3_{S}:=(\ldots)_1(\ldots)_2(\ldots)_3 \,,
\een
where the sub-indices $1,2,3$ refer to  the following identification of the parameters  $e_1,e_2,e_3$ to the squashing parameters $\omega_1, \omega_2,\omega_3$
in each sector:
\ben\label{p123}
\begin{array}{|c|ccc|}
\hline
&\quad e_1\quad &\quad e_2\quad &\quad e_3\quad\\
\hline
1&\omega_3&\omega_2&\omega_1\\
2&\omega_1&\omega_3&\omega_2\\
3&\omega_1&\omega_2&\omega_3\\
\hline
\end{array}
\een
Explicitly we have: 
\ben\nn
\left({\cal Z}_{\rm inst}\right)_1=
{\cal Z}^{\mathbb{R}^4\times S^1}_{\rm inst}\left(\frac{i \sigma}{\omega_1}, \frac{\vec m}{\omega_1};\frac{\omega_2}{\omega_1},\frac{\omega_3}{\omega_1}\right)\,,  &&\,
\left({\cal Z}_{\rm inst}\right)_2=
{\cal Z}^{\mathbb{R}^4\times S^1}_{\rm inst}\left(\frac{i \sigma}{\omega_2}, \frac{ \vec m}{\omega_2};\frac{\omega_1}{\omega_2},\frac{\omega_3}{\omega_2}\right)\, ,\\
&& \!\!\!\!\!\!\!\!\!\!\! \!\!\!\!\!\!\!\!  \!\!\!\!\!\!\!\!\!\!\! \!\!\!\!\!\!\!\! 
\!\!\!\!\!\!\!\!\!\!\! \left({\cal Z}_{\rm inst}\right)_3=
{\cal Z}^{\mathbb{R}^4\times S^1}_{\rm inst}\left(\frac{i \sigma}{\omega_3}, \frac{ \vec m}{\omega_3};\frac{\omega_1}{\omega_3} ,\frac{\omega_2}{\omega_3}\right)\,.
 \een
Notice that since the instanton partition function can be expressed in terms of the 
parameters   
\be\label{pq}
q=e^{2\pi i e_1/e_3}\,,\quad t=e^{2\pi ie_2/e_3}\,,
\ee
the  pairing   defined in eq. (\ref{sl3}) enjoys an $SL(3,\mathbb{Z} )$ symmetry
which acts {\it S-dualizing} the  couplings $q$ and $t$.\\

{The classical factor contains the contribution of the Yang-Mills action  given by \footnote{To simplify formulas we define $ g^2= \frac{g_{YM}^2}{4i \pi^2}$.}
\be\label{class}
{Z}_{\rm cl}(\sigma)=
e^{\frac{2\pi i}{\omega_1\omega_2\omega_3 g^2}\,\text{Tr}(\sigma^2)}=
e^{-\frac{2\pi i}{\omega_1\omega_2\omega_3 g^2 2C_2(ad)} \sum_\alpha [i\alpha(\sigma)]^2}
\ee
where we denoted by $\alpha$ a root of the Lie algebra of the gauge group
and we used that   $2C_2(ad)\sum_\rho\rho(\sigma)^2=\sum_\alpha \alpha(\sigma)^2$. 
We can try to bring this term in the $SL(3,\mathbb{Z})$ factorized form as the instanton contribution. We begin by rewriting the classical term
by means of the Bernoulli polynomial $B_{33}$, defined in appendix \ref{ber}, as
\be
{Z}_{\rm cl}(\sigma)= \prod_\alpha e^{-\frac{2\pi i}{3!}\left[B_{33}\left(i\alpha(\sigma)+\frac{1}{g^22C_2(ad)}+\frac{E}{2}\right)-B_{33}\left(\frac{1}{g^22C_2(ad)}+\frac{E}{2}\right)\right]}.
\ee
Each  factor in the above expression can in turn be factorized thanks to the  following identity  \cite{fv}
\ben\label{nfac}
e^{-\frac{2\pi i}{3!}B_{33}(z)}&=& \prod_{k=1}^3\Gamma_{q,t}\left(\frac{z}{e_3}\right)_{k}=\Big|\Big|  \Gamma_{q,t}\left(\frac{z}{e_3}\right) \Big|\Big|^3_{S}  \,,
\een
where  the elliptic gamma function $\Gamma_{q,t}$ is defined in appendix \ref{tega}.
Hence by applying  the identity (\ref{nfac}) to each Bernoulli factor we  obtain:\footnote{A Chern-Simons term can be similarly dealt with by writing cubic terms  as sums of
$B_{33}$.}
\ben\label{clfac}
{Z}_{\rm cl}(\sigma)=\Big|\Big|  {\cal Z}_{\rm cl}\Big|\Big|^3_{S} \, ,
\een
with 
\ben
\label{zcl}
{\cal Z}_{\rm cl}=\prod_\alpha \frac{\Gamma_{q,t}\left(\frac{1}{e_3}\left(i\alpha(\sigma)+\frac{1}{g^22C_2(ad)}+
\frac{E}{2}\right)\right)}{\Gamma_{q,t}\left(\frac{1}{e_3}\left(\frac{1}{g^22C_2(ad)}+\frac{E}{2}\right)\right)}\,.
\een}
We can therefore write the  partition function as:
\ben
\label{good2cft}
Z_{S^5}= \int~d \sigma~ { Z}_{\text{1-loop}}(\sigma,\vec M)~ 
 \Big|\Big| {\cal F}\Big|\Big|^3_{S}\,,
\een
with
\be
\label{fifi}
 {\cal F}={\cal Z}_{\rm cl} \, {\cal Z}_\text{inst}\,,
\ee
where ${\cal Z}_{\rm cl} $ and ${\cal Z}_\text{inst}$ are given respectively in (\ref{zcl}) and (\ref{cinst}).

We can give another representation of the partition function where we bring 
also the 1-loop contribution to an $SL(3,\mathbb{Z})$ factorized form as in  \cite{Lockhart:2012vp}.
We remind that a vector multiplet  contributes to the partition function with
\be
{ Z}^{\text{vect}}_{\text{1-loop}}(\sigma)=\prod_{\alpha>0} S_3(i\alpha(\sigma)) S_3(E+i\alpha(\sigma))\,,
\ee
while a {hyper multiplet} of mass $M$ and representation $R$ gives
\be
{ Z}^{\text{hyper}}_{\text{1-loop}}(\sigma,M,R)=\prod_{\rho\in R} S_3\left({i\rho(\sigma)}+iM+\frac{E}{2}\right)^{-1}\,,
\ee
where  $S_3$ is the triple sine function defined in appendix \ref{tri}. Using the relation (\ref{s3fac}),  the vector multiplet contribution can be written as 
\ben
{ Z}^{\text{vect}}_{\text{1-loop}}(\sigma)&=&\prod_{\alpha>0} e^{-\frac{\pi i }{3!}{[B_{33}(i\alpha(\sigma))+B_{33}(i\alpha(\sigma)+E)]}} 
(1-e^{\frac{2\pi i}{\omega_1}[i\alpha(\sigma)]})
(1-e^{\frac{2\pi i}{\omega_2}[i\alpha(\sigma)]})(1-e^{\frac{2\pi i}{\omega_3}[i\alpha(\sigma)]})
\nn\\&&
\times\,
\prod_{k=1}^3  (te^{\frac{2\pi i}{e_3}[i\alpha(\sigma)]};q,t)_k(qe^{\frac{2\pi i}{e_3}[i\alpha(\sigma)]};q,t)_k\,,
\een 
where $(z;q,t)=\prod_{i,j\geq 0}(1-zq^i t^j)$ denotes the double $(q,t)$-factorial, while the sub-index $k$ refers to the way $q,t$ defined in (\ref{pq}) are related to  the squashing parameters according to  (\ref{p123}).
Each factor
$(te^{\frac{2\pi i}{e_3}[i\alpha(\sigma)]};q,t)_k(qe^{\frac{2\pi i}{e_3}[i\alpha(\sigma)]};q,t)_k$
can in turn be identified with the 1-loop  vector multiplet contribution to  the ${\mathbb R}^4\times S^1$ theory 
\cite{Nekrasov:2002qd, Nekrasov:2003rj, Alday:2009aq} appropriately rescaled, and with  equivariant parameters 
as in  (\ref{pq}).
We can equivalently factorize the vector multiplet {1}-loop term in the following more compact form:
{\ben
\label{ff1}
 Z^{\text{vect}}_{\text{1-loop}}(\sigma)&=&\prod_{\alpha}e^{-\frac{\pi i}{3!}B_{33}(i\alpha(\sigma))}\prod_{k=1}^3(e^{\frac{2\pi i}{e_3} {[}i\alpha(\sigma){]}};q,t)_k.
\een}
Analogously, the {hyper multiplet} factor can be written as 
\ben
\label{ff2}
{Z}^{\text{hyper}}_{\text{1-loop}}(\sigma,M,R)&=&\prod_{\rho\in R} e^{\frac{\pi i }{3!}
B_{33}(i\rho(\sigma)+iM+\frac{E}{2})} \prod_{k=1}^3 
 (e^{\frac{2\pi i}{e_3} [ i\rho(\sigma)+i M+\frac{E}{2} {]} };q,t)_k^{-1}\,,
\een
where $(e^{\frac{2\pi i}{e_3} {[}i\rho(\sigma)+i M+\frac{E}{2} {]} };q,t)^{-1}$ can be identified with the
1-loop {hyper multiplet} contribution to  the ${\mathbb R}^4\times S^1$
partition function \cite{Nekrasov:2002qd, Nekrasov:2003rj, Alday:2009aq}. 
The full 1-loop contribution to the partition function as can be therefore written as
\be
Z_{\text{1-loop}}(\sigma,\vec M)=\prod_R\prod_{\substack{\alpha\\
\rho\in R}}e^{-\frac{\pi i}{3!}[B_{33}(i\alpha(\sigma))-B_{33}(i\rho(\sigma)+m_R)]}\prod_{k=1}^3\frac{(e^{\frac{2\pi i}{e_3} [i\alpha(\sigma)]};q,t)_k}{(e^{\frac{2\pi i}{e_3}[i\rho(\sigma)+m_R]};q,t)_k}
\ee
where  $m_R=iM_R+E/2$ is the mass of a multiplet in the representation $R$. If we consider (pseudo) real representations, for each weight $\rho$ there is the opposite weight $-\rho$, in which case the Bernoulli sum to a quadratic polynomial which can be easily factorized in terms of elliptic Gamma functions as we did for the classical term. In fact, in this case the Bernoulli from the 1-loop factor will amounts to a renormalization of the gauge coupling constant. To see this, let us consider $N_f$ fundamentals of mass $M_f$ and  $N_f$ anti-fundamentals of mass $\bar M_f$, with $f=1,\ldots,N_f$, and $N_a$ adjoints of mass $M_a$, $a=1,\ldots, N_a$. The total contribution from 1-loop Bernoulli terms is (up to a $\sigma$-independent constant)
\be
\label{xox}
e^{-\frac{2\pi i}{\omega_1\omega_2\omega_3} \frac{\sum_\alpha[i\alpha(\sigma)]^2}{2C_2(ad)}\left[\frac{E}{4}N_f-\frac{1}{4}\sum_f(m_f+\bar m_f)+C_2(ad)(\frac{E}{2}(N_a-1)-\sum_a m_a)\right]}\,,
\ee
since  $\sum_\rho\rho(\sigma)=0$ for $SU(N)$. Comparing the above expression with
the classical action (\ref{class}), it is easy to obtain  the factorized version  of (\ref{xox})
by  shifting the gauge coupling constant in (\ref{zcl}), according to
\be
\frac{1}{g^2}\rightarrow \frac{1}{g^2}+\frac{E}{4}N_f-\frac{1}{4}\sum_f(m_f+\bar m_f)+C_2(ad)\left(\frac{E}{2}(N_a-1)-\sum_a m_a\right)\,.
\ee
Therefore also the total 1-loop contribution admits a factorized form:
\ben\label{clfac1}
{Z}_{\rm 1-loop}(\sigma,\vec{M})=\Big|\Big|  {\cal Z}_{\text{1-loop}}\Big|\Big|^3_{S} \, .
\een
Putting  all together  we can finally write (up to constant prefactors):
\be
Z_{S^5}=\int d \sigma ~\Big|\Big| {\cal  B}^{5d} \Big|\Big|^3_{S}\,,
\ee
where  ${\cal  B}^{5d}$, the  5d holomorphic block, is defined as
\be
\label{5hb}
{\cal  B}^{5d}= {\cal  Z}_{\text{1-loop}}~{\cal  Z}_{\text{cl}}  ~{\cal Z}_{\text{inst}}\,.
\ee
It is important to note the way we factorize 1-loop and classical terms is not  unique.
We have seen an example of this for the vector multiplet contribution. We will now  see more precisely how to track this ambiguity in an example.

\subsubsection*{The $SU(2)$ superconformal QCD}

We consider the superconformal QCD (i.e. SCQCD) with $SU(2)$ gauge group.
In this case, the Coulomb branch parameter can be written as 
$\sigma=i\bigl(\begin{smallmatrix}a_1&0\\ 0&a_2\end{smallmatrix} \bigr)$ with $a_1=-a_2=a$ 
and the {vector multiplet}
is coupled to four fundamental {hyper multiplets} with masses {$M_f$, $f=1,2,3,4$.}
The total {1}-loop contribution is given by:
\be
\label{scqcd1loop}
Z_{\rm 1-loop}(a,\vec M)  = \frac{S_3\left( a_1-a_2\right)S_3\left( a_2-a_1\right)}
{\prod_{f} S_3\left( a_1+m_f\right)S_3\left( a_2+m_f\right)}\,.
\ee
Collecting  the $B_{33}$ factors coming from the factorization of the $S_3$ functions  in (\ref{scqcd1loop}) we find:
\be
B_{33}(2a)+B_{33}(-2a)-\sum_{f} [B_{33}(a+m_f)+B_{33}(-a+m_f)]=
-\frac{6a^2}{\omega_1\omega_2\omega_3}\sum_{f}m_f+const\,,
\ee
where the constant {denotes $a$-independent terms}. We then combine {1-}loop and the  classical Yang-Mills action
$$
Z_{\text{cl}}(a)=e^{-\frac{2\pi i}{\omega_1\omega_2\omega_3}\frac{a^2}{\tilde g^2}}\,,
$$
where  $\tilde g^2=g^2/2$ obtaining: 
\be
Z_{\text{cl}}(a)~ {Z}_{\text{1-loop}}(a,\vec M)  = e^{-\frac{2\pi i a^2}{\omega_1\omega_2\omega_3}\left(\frac{1}{\tilde g^2}-\frac{1}{2}\sum_f m_f\right)}  
\prod_{k=1}^3\frac{(e^{\frac{2\pi i}{e_3}[\pm 2a]};q,t)_k}{\prod_{f=1}^4(e^{\frac{2\pi i}{e_3}[\pm a+m_f]};q,t)_k}\,,
\ee
where for compactness we have employed the shorthand notation $f(\pm a)=f(a)f(-a)$. The next step is to write  the exponential as a sum of $B_{33}$,
\be
e^{-\frac{2\pi i a^2}{\omega_1\omega_2\omega_3}\left(\frac{1}{\tilde g^2}-\frac{1}{2}\sum_f m_f\right)} =
\frac{e^{-\frac{2\pi i}{3!}B_{33}(\pm a+\frac{1}{\tilde g^2}-\frac{1}{2}\sum_f m_f+\kappa)}}{e^{-\frac{2\pi i}{3!}B_{33}(\pm a+\kappa)}}
\times
\frac{e^{-\frac{4\pi i}{3!}B_{33}(\kappa)}}{e^{-\frac{4\pi i}{3!}B_{33}(\frac{1}{\tilde g^2}-\frac{1}{2}\sum_fm_f+\kappa)}}\,.
\label{cane}
\ee
In this expression  the coefficient $\kappa$ appearing only on the R.H.S.  is completely  arbitrary and can be identified
with the ambiguity of the factorization. 
Finally we apply the identity (\ref{nfac}) to each Bernoulli factor in (\ref{cane})
and obtain the SQCD 5d holomorphic blocks:\footnote{We are dropping $a$-independent elliptic Gamma factors.}
\be
\mathcal{B}^{\rm 5d}= 
\frac{(e^{\frac{2\pi i}{e_3}[\pm 2a]};q,t)}{\prod_{f}(e^{\frac{2\pi i}{e_3}[\pm a+m_f]};q,t)}\cdot
\frac{\Gamma_{q,t}\left(\frac{\pm a+1/\tilde g^2-\sum_f m_f/2+\kappa}{e_3}\right)}
{\Gamma_{q,t}\left(\frac{\pm a +\kappa}{e_3}\right)}\cdot
\mathcal{Z}_\text{inst}\,.\ee

\subsection{$S^4\times S^1$ partition functions and 5d holomorphic blocks}

The partition functions for 5d ${\cal N}=1$ supersymmetric gauge theory on  $S^4\times S^1$ has been  computed in \cite{Kim:2012gu,Terashima:2012ra,Iqbal:2012xm} and reads
\ben
\label{s1s4}
Z_{S^4\times S^1}&=&\int d\sigma ~ {Z}_{\text{1-loop}}(\sigma,\vec M)~  { Z}_{\text{inst}}(\sigma,\vec M) \,,
\een
where $\sigma$ is the Coulomb branch parameter.
The instanton part  receives contributions   from the fixed points at north and south poles of the
 $S^4$ and  can be written as 
\be
 Z^{}_\text{inst}=\Big|\Big|{\cal Z}_\text{inst}\Big|\Big|^2_{id} \,,
\ee
where  as before ${\cal Z}^{}_{\rm inst}$ coincides with  the equivariant instanton partition function 
on ${\mathbb R}^4\times S^1$  with Coulomb and mass parameters appropriately 
rescaled, and a particular parameterization of the equivariant parameters:
\be 
 {\cal Z}^{}_\text{inst}(\sigma,\vec m)=  {\cal Z}^{\mathbb{R}^4\times S^1}_\text{inst} \left(\frac{i \sigma}{e_3}, \frac{\vec m}{e_3}; 
 \frac{e_1}{e_3},\frac{e_2}{e_3}\right)\,,
\ee
with $m_j=i M_j + Q_0/2$ and $Q_0=b_0+1/b_0$.  We also introduced  the  5d $id$-pairing  defined as
\ben\label{id}
\Big|\Big| (\dots)\Big|\Big|^2_{id}:=(\ldots)_1(\ldots)_2\,,
\een
where the 1,2 sub-indices means that the  $e_1,e_2,e_3$ parameters assume the following values:
\ben\label{idqt}
\begin{array}{|c|ccc|}
\hline
&\quad e_1\quad&\quad e_2\quad&\quad e_3\quad\\
\hline
1&b_0^{-1}&b_0&2\pi i/\beta\\
2&b_0^{-1}&b_0&-2\pi i/\beta\\
\hline
\end{array}\nn
\een
with  $\beta$  the circumference  of $S^1$ and  $b_0$   the squashing parameter  of  $S^4$.

Due to the property   (\ref{gamma unit})  the elliptic Gamma function satisfies $\Big|\Big| \Gamma_{q,t}(z)\Big|\Big|^2_{id}=1$  and
 the classical term  ${\cal  Z}_{\text{cl}}$ defined in eq. (\ref{zcl})
{``squares"} to one:
\be
\Big|\Big| {\cal  Z}_{\text{cl}}\Big|\Big|_{id}^2=1\,.
\ee 
We can therefore   write 
\ben
\label{idcft2}
Z_{S^4\times S^1}&=&\int d \sigma ~{ Z}_{\text{1-loop}}(\sigma,\vec M) ~ \Big|\Big| {\cal F}\Big|\Big|^2_{id}\,,
\een
with ${\cal F}$  defined in eq. (\ref{fifi}). 

The {1}-loop contributions of vector   and hyper multiplets are given respectively by
\be
{ Z}^{\text{vect}}_{\text{1-loop}}(\sigma)=\prod_{\alpha>0}\Upsilon^\beta\left(i\alpha(\sigma)\right)\Upsilon^\beta\left(-i\alpha(\sigma)\right)\,,
\ee
and
\be
{ Z}^{\text{hyper}}_{\text{1-loop}}(\sigma,M,R)=\prod_{\rho\in R}\Upsilon^\beta\left({i\rho(\sigma)+iM}+
\frac{Q_0}{2}\right)^{-1}\,,
\ee
where $\Upsilon^\beta$  is defined in appendix \ref{upb}.
Also in this case it is possible to bring the {1}-loop term in a factorized form. Indeed 
if we use again that $\Big|\Big| \Gamma_{q,t}(z)\Big|\Big|^2_{id}=1$ we can write
\be
{ Z}^{\text{vec}}_{\text{1-loop}}(\sigma) =\Big|\Big| {\cal Z}^{\text{vec}}_{\text{1-loop}}\Big|\Big|^2_{id}\,,
\qquad 
{ Z}^{\text{hyper}}_{\text{1-loop}}(\sigma,\vec M) =\Big|\Big| {\cal Z}^{\text{hyper}}_{\text{1-loop}}\Big|\Big|^2_{id}\,,
\ee
with  ${\cal Z}^{\text{vec}}_{\text{1-loop}},{\cal Z}^{\text{hyper}}_{\text{1-loop}}$
coinciding  with the corresponding  factors obtained from the factorization   of {1}-loop terms on $S^5$, which implies that
\be
Z_{S^4\times S^1}=\int d \sigma  ~\Big|\Big| {\cal  B}^{5d}\Big|\Big|^2_{id}\,.
\ee
Hence we  surprisingly  discover that for a given theory, the $S^5$ and  $S^4\times S^1$ partition functions 
can be factorized in terms of the same 5d blocks  ${\cal  B}^{5d}$.
This is very reminiscent to what happens in 3d, where the partition function for $S^3_b$ and $S^2\times S^1$ can be expressed via the same blocks, paired in two different ways \cite{hb}.

It would be very interesting to test whether partition functions on more general 5-manifolds
can be engineered by fusing our ${\cal  B}^{5d}$ blocks with suitable gluings. 

\subsection{Degeneration of $5d$ partition functions}\label{secdeg}

An interesting feature of  $S^5$ and $S^4\times S^1$ partition functions (\ref{s5}), (\ref{s1s4}) 
is that for particular values of the  masses 1-loop factors develop poles which  pinch the integration 
contour. The partition functions can then be defined via a meromorphic analytic continuation which prescribes
to take the residues at the  poles  trapped along the integration path.
Here we will consider a particular example, the $SU(2)$ theory
with four fundamental hyper multiplets with masses $M_f$,  $f=1,\ldots ,4$ on $S^5$. In this case when two of the masses, say $M_1, M_2$, satisfy the following condition:
\be
M_1+M_2=i (\omega_3+E) \qquad {\rm or}  \qquad m_1+m_2=-\omega_3\,,
\label{deg1}
\ee
where $m_i$ are defined in (\ref{newmass}),
the partition function receives contribution only from two pinched  poles located   at 
\be
a_1=m_{1}=-m_2-\omega_3=-a_2\,, \qquad a_1=m_{1}+\omega_3=-m_2=-a_2\,.
\label{pol1}
\ee 
This  is reminiscent of the degeneration of a Liouville theory 4-point correlation function  
when one of the momenta is analytically continued to a degenerate value.
The  AGT setup explains that the degeneration limit corresponds on the gauge theory side  to   the reduction  of   4d partition functions to  simple surface operator partition functions  \cite{Alday:2009fs, Dimofte:2010tz,Kozcaz:2010af,Bonelli:2011fq}. In particular the $S^4$ SCQCD partition function, in this limit, reduces  to its codimension-2 BPS defect theory,  the $S^2$  SQED Higgs branch partition function  \cite{Doroud:2012xw,Benini:2012ui}.
Therefore we expect that, when the masses are analytically continued to the values (\ref{deg1}), the squashed $S^5$  SCQCD partition function  will degenerate to its codimension-2 defect theory, the 3d SQED  defined on  the squashed 3-sphere $S_b^3$ which we record below.
The $S^3_b$  Higgs branch  partition function of the $U(1)$, $\mathcal{N}=2$ theory with 2 charge plus and 2 charge minus chirals,  
takes the following form \cite{Pasquetti:2011fj}:
\be\label{Zellipsoid}
Z_{S^3_b}= \sum_{i=1,2} G^{(i)}_{\rm cl}  G^{(i)}_{\rm 1-loop} \Big|\Big|\mathcal{Z}^{(i)}_V\Big|\Big|^2_S\,,
\ee
where the sum runs  over the two SUSY vacua of theory  and   
\be
\label{cl3p}
G^{(i)}_{\rm cl}=e^{-2\pi i \xi m^{3d}_i}
\,, 
\qquad G^{(i)}_{\rm 1-loop}=\prod_{j,k=1,2} \frac{s_b(m^{3d}_j-m^{3d}_i+i Q/2)}{s_b(\tilde m^{3d}_k-m^{3d}_i-i Q/2)}\,,\quad j\neq i\, ,
\ee
\ben
\mathcal{Z}^{(i)}_V=\sum_{n\geq 0}\prod_{j,k=1,2} \frac{( y_k x_i^{-1};q)_n}{(q x_j x_i^{-1};q)_n}  (u^{3d})^n\, ,
\een
where $b$ is the squashing parameter of the ellipsoid, $Q=b+1/b$, and $s_b$ is the double sine function defined in appendix \ref{ber}.
The vortex partition functions $\mathcal{Z}^{(i)}_V$  are basic hypergeometric functions defined in (\ref{qhyper21}) with coefficients: 
\ben
\label{etil}\nn
x_i&=&e^{2\pi b m^{3d}_i}, \quad y_i=e^{2\pi b \tilde m^{3d}_i}, \quad  z^{3d}=e^{2\pi b \xi}, \quad q=e^{2\pi i b^2}\,,\\
\tilde x_i&=&e^{2\pi  m^{3d}_i/b}, \quad \tilde y_i=e^{2\pi  \tilde m^{3d}_i/b}, \quad \tilde  z^{3d}=e^{2\pi \xi/b},
 \quad \tilde q=e^{2\pi i/ b^2} \,,
\een
and
\be \prod_{j,k=1,2}x_jy_k^{-1}=r,\quad  (u^{3d}) = q r^\frac{1}{2} (z^{3d})^{-1}\,.
\ee
Finally the 3d $S$-pairing is defined as \cite{hb}:
\be
\label{sp}
\Big| \Big| f(x;q)\Big| \Big|_{S}^2=f(x;q)f(\tilde x;\tilde q)\,.
\ee
We will now show how to reconstruct (\ref{Zellipsoid})  from the residues of the $S^5$ partition function at the pinched poles.
We begin by analyzing  the SCQCD  instanton contribution which  is given by:
\ben\label{nek}
 {\cal Z}_{\rm inst}=  {\cal Z}^{\mathbb{R}^4\times S^1}_{\rm inst}\left(\frac{\vec{a}}{e_3},
\frac{\vec{m}}{e_3} ;  \frac{e_1}{e_3},\frac{e_2}{e_3}\right)=\sum_{\vec{Y}} z^{|\vec{Y}|}\frac{F_{\vec{Y}}(\vec{a},\vec{m})}{V_{\vec{Y}}(\vec{a})}\,,
\qquad z=e^{\frac{2\pi i}{ e_3  \tilde g^2 }}\,.
\een
The numerator $F_{\vec{Y}}(\vec{a},\vec{m})$ encodes the contribution of the four hyper multiplets and the denominator $V_{\vec{Y}}(\vec{a})$ is due to the vector multiplet and we refer to  appendix \ref{instdeg} for explicit expressions. 
As explained in appendix \ref{instdeg}, when the Coulomb branch parameters take  the values  $a_1=m_{1}=-m_2-\omega_3=-a_2$, the  instanton partition function (\ref{nek})  degenerates to a basic  hypergeometric function.
In particular the  instanton partition function in the first sector becomes:
\ben
\left({\cal Z}_{\rm inst}\right)_1={\cal Z}^{\mathbb{R}^4\times S^1}_{\rm inst}\left(\frac{\vec{a}}{\omega_1},\frac{\vec{m}}{\omega_1};\frac{\omega_2}{\omega_1},\frac{\omega_3}{\omega_1}\right)\
\quad\xrightarrow[(a_1,a_2)\rightarrow(m_{1},m_2+\omega_3)]{}\quad\phantom{|}_2\Phi_1(A,B;C,e^{2\pi i\frac{\omega_2}{\omega_1}};u)\,,\nonumber\\
\een
where $A,B,C,u$ parameters are defined in (\ref{par}) and we replaced $e_1,e_2,e_3$ in eq. (\ref{hyperdeg}) with their 
values in terms of $\omega_i$ in sector 1 as in (\ref{p123}).
In sector 2 we use (\ref{hyperdeg2}) and the appropriate values of $e_i$ in terms of $\omega_i$
to find:
\ben
\left({\cal Z}_{\rm inst}\right)_2={\cal Z}^{\mathbb{R}^4\times S^1}_{\rm inst}
\left(\frac{\vec{a}}{\omega_2},\frac{\vec{m}}{\omega_2};\frac{\omega_1}{\omega_2},\frac{\omega_3}{\omega_2}\right)
\quad\xrightarrow[(a_1,a_2)\rightarrow(m_{1},m_2+\omega_3)]{}
\quad\phantom{|}_2\Phi_1(\tilde A,\tilde B;\tilde C,e^{2\pi i\frac{\omega_1}{\omega_2}};\tilde u)\,,\nonumber\\
\een
where the tilde symbol  indicates  $\omega_1 \leftrightarrow \omega_2$.  
Finally in   sector 3, once the $e_i$ are expressed in terms of the $\omega_i$ according to (\ref{p123}), 
all the parameters are  rescaled by  $\omega_3$ yielding a trivial degeneration, as explained  in appendix \ref{instdeg}:
\ben
\left({\cal Z}_{\rm inst}\right)_3=
{\cal Z}^{\mathbb{R}^4\times S^1}_{\rm inst}\left(\frac{\vec{a}}{\omega_3},\frac{\vec{m}}{\omega_3};\frac{\omega_1}{\omega_3},
\frac{\omega_2}{\omega_3}\right)\quad\xrightarrow[(a_1,a_2)\rightarrow(m_{1},m_2+\omega_3)]{}\quad
1\,.
\een
Putting all together we find that
\ben
\Big|\Big| {\cal Z}_{\rm inst}\Big|\Big|^3_{S}\quad\xrightarrow[(a_1,a_2)\rightarrow(m_{1},m_2+\omega_3)]{}\quad 
\Big|\Big| \phantom{|}_2\Phi_1(A,B;C,e^{2\pi i\frac{\omega_2}{\omega_1}};u)   
\Big|\Big|^2_S=\Big|\Big| {\cal Z}^{(1)}_V\Big|\Big|^2_S\,,
\label{gat}
\een
which shows that remarkably the 5d $S$-pairing  reduces to the 3d  $S$-pairing.
Furthermore by identifying the coefficient $A,B,C$ of the basic hypergeometrc function with  those of the vortex partition
function $\mathcal{Z}^{(1)}_V$,  as required by  last equality in eq. (\ref{gat}),  we obtain
the following dictionary between 3d and 5d masses:
\be
\label{53dictio}
m^{3d}_1=-im_1\,,\quad m^{3d}_2=-im_2\,,\quad \tilde m^{3d}_1=im_3\,,\quad \tilde m^{3d}_2=im_4\,,
\ee
while by matching the expansion parameters we find
\be
\label{53dic2}
i\xi = 1/\tilde g^2\,.
\ee
Finally, we also identify
\be
\omega_2=\frac{1}{\omega_1}=b\,.
\ee
In complete analogy for the other pole, located  at 
$a_1=m_{1}+\omega_3=-m_2=-a_2$,  upon using  the dictionary (\ref{53dictio}), we find
\ben
\Big|\Big| {\cal Z}_{\rm inst}\Big|\Big|^3_{S}\quad\xrightarrow[(a_1,a_2)\rightarrow(m_{1}+\omega_3,m_2)]{}\quad 
\Big|\Big|\mathcal{Z}^{(2)}_V
\Big|\Big|^2_S\,.
\een
We next consider the 1-loop term given in (\ref{scqcd1loop}).
Since this term  has   poles when $(a_1,a_2)$ take the two  values in (\ref{pol1}) and since we are only
 interested in showing that the degeneration of the
$S^5$ partition function reproduces   the $S^3_b$ partition function  up to a prefactor, 
we  evaluate   the  ratio of the residues at each pole. This ratio is finite and, by using the property (\ref{prorat}), it  is straightforward to show that it
 reproduces the ratio of the  $S^3_b$
1-loop terms in the two SUSY vacua, given in eq. (\ref{cl3p}):
\be
\frac{Z_{\rm 1-loop}\Big|_{(a_1,a_2)\rightarrow(m_{1},m_2+\omega_3)}}
{Z_{\rm 1-loop}\Big|_{(a_1,a_2)\rightarrow(m_{1}+\omega_3,m_2)}}=\frac{G^{(1)}_{\rm 1-loop}}{G^{(2)}_{\rm 1-loop}}\,.
\ee
 Similarly, it is simple to show that the ratio of the residues of the classical term at the points (\ref{pol1})
reproduces the ratio of the $S^3_b$
classical  terms in the two SUSY vacua, given in eq. (\ref{cl3p}):
\be
\frac{Z_{\rm cl}\Big|_{(a_1,a_2)\rightarrow(m_{1},m_2+\omega_3)}}
{Z_{\rm cl}\Big|_{(a_1,a_2)\rightarrow(m_{1}+\omega_3,m_2)}}=
e^{-\frac{2\pi i }{\omega_1\omega_2\omega_3}\frac{(m_1^2-m_2^2)}{\tilde g^2}}
=e^{\frac{2\pi i (m_1-m_2) }{\omega_1\omega_2 \tilde g^2}
}=e^{2\pi i\xi(m_1^{\rm 3d}-m_2^{\rm 3d})}=\frac{G^{(1)}_{\rm cl}}{G^{(2)}_{\rm cl}}\,,
\ee
where we used  $m_1^2-m_2^2
=-\omega_3 (m_1-m_2)$ and the dictionary  (\ref{53dictio}), (\ref{53dic2}).

Finally putting all together we obtain the promised result:
\be
Z^{SCQCD}_{S^5}
\quad\xrightarrow[m_{1}+m_2=-\omega_3]{}\quad 
 \sum_i ^2 G^{(i)}_{\rm cl}G^{(i)}_{\rm 1-loop}\Big|\Big| {\cal Z}^{(i)}_V\Big|\Big|^2_S=Z^{SQED}_{S^3}\,.
\ee
Notice that there are two extra choices for the degeneration condition, which would have
led to the same result: 
\ben
\nonumber
&& m_1+m_2=-\omega_1\,, \quad {\rm with } \quad \omega_2=\frac{1}{\omega_3}=b\,,\\
&& m_1+m_2=-\omega_2\,,  \quad {\rm with } \quad \omega_1=\frac{1}{\omega_3}=b\,. 
\een
The three possibilities  correspond to choosing one of the three maximal squashed 3-spheres inside the squashed 5-sphere.

In a similar manner, it is possible to show that the partition function of the SCQCD on $S^4\times S^1$, 
when two of the masses satisfy the condition
\be
m_1+m_2=-b_0\,,
\label{pol2}
\ee 
reduces to the SQED partition function on $S^2\times S^1$:\footnote{The  explicit expression of the Higgs branch partition function  $Z^{SQED}_{S^2\times S^1}$ can be found in  \cite{Nieri:2013yra}. }
\be
Z^{SCQCD}_{S^4\times S^1}
\quad\xrightarrow[m_{1}+m_2=-b_0]{}\quad 
 \sum_i ^2 G^{(i)}_{\rm cl}G^{(i)}_{\rm 1-loop}\Big|\Big| {\cal Z}^{(i)}_V\Big|\Big|^2_{id}=Z^{SQED}_{S^2\times S^1}\,,
\ee
with  the 3d angular momentum fugacity $q$ related to the 5d parameters by
 $q=e^{\beta/b_0}$.
Also in this case there is another possible  degeneration condition $m_1+m_2=-\frac{1}{b_0}$, 
which leads to the same result but with the identification  $q=e^{\beta b_0}$.
The two choices, correspond to the two  maximal $S^2$ inside the squashed $S^4$.

\subsubsection{Higher degenerations}\label{sechdeg}

In the previous section we focused on particular analytic continuations of the 5d masses
(\ref{pol1}), (\ref{pol2}) which degenerate partition functions to a bilinear combination
of solutions to  basic  hypergeometric difference equations.
The degeneration mechanism is actually much more general, for example the 1-loop factor
of the SCQCD on $S^5$  develops poles pinching the integration contour when two of the masses
satisfy the following condition:
\be
\label{highermas}
m_1+m_2=-(n_1\omega_1+n_2\omega_2+n_3\omega_3)=-\vec n \cdot \vec \omega \,, \qquad n_1,n_2,n_3\in \mathbb{Z}_+\,.
\ee
In this case the integral localizes to a sum over the following set:
\be
\{a^*\}: \quad a_1=m_1+ (\vec n-\vec p)\cdot \vec \omega \,,\quad a_2=m_2+\vec p \cdot \vec \omega \,,
\label{pol3}
\ee
where
\be
p_k\in \{0,1, \cdots , n_k\}\,, \quad k=1,2,3\,.
\ee
Evaluating the residues at points (\ref{pol3}) we find:
\ben
\label{to3}
Z^{SCQCD}_{S^5}
\quad\xrightarrow[m_1+m_2=-\vec n \cdot \vec \omega]{}\quad 
\sum_{\{ a^*\}} 
\text{ Res}\Big[Z_{\rm 1-loop} \Big] ~~({\cal F}^{p_2,p_3})_1 ~({\cal F}^{p_3,p_1})_2~( {\cal F}^{p_2,p_1})_3\,,
\een
where $({\cal F}^{p_i,p_j})$ denotes the value  of classical and instanton parts at the pole
\be
({\cal F}^{p_i,p_j})_k=({\cal Z}^{p_i,p_j}_{\rm cl})_k  ({\cal Z}^{p_i,p_j}_{\rm inst})_k\,, \quad {\rm with} \quad i\neq j\neq k=1,2,3\,,
\ee
and, as  usual,  the subscript $k$ means that in each sector we express the $e_i$ in terms of the $\omega_i$s according to the dictionary (\ref{p123}).

The residue of the instanton contribution is given by:
\be
{\cal Z}^{p_2,p_1}_{\rm inst}=\sum_{\substack{Y^1=(p_2,\,p_1)\\
Y^2= (n_2-p_2,\,n_1-p_1)}} \frac{F_{p_2,p_1}}{ V_{p_2,p_1}} ~~z^{|\vec Y|}\,,
\ee
where $F_{p_2,p_1}$ and  $V_{p_2,p_1}$ are defined in eqs. (\ref{morte1}) and (\ref{morte2}).
In this case the analytic continuation of the masses restricts the sum to hook Young tableaux with $p_2$ rows and $p_1$ columns, and with $n_2-p_2$ rows and $n_1-p_1$ columns respectively.
Furthermore ${\cal Z}^{p_3,p_2}_{\rm inst}$ and ${\cal Z}^{p_3,p_1}_{\rm inst}$
are obtained from ${\cal Z}^{p_2,p_1}_{\rm inst}$ by appropriately renaming/permuting
the $p_i$.

The value of the classical term ${\cal Z}^{p_2,p_1}_{\rm cl}$ is computed in eqs. (\ref{degcl1}), (\ref{degcl2}) and the identification of the  $p_i$ as well as of the $e_i$ is like in the instanton case discussed above.

In the next section we will provide an interpretation of the degeneration (\ref{to3}) using the correspondence between the $S^5$ SCQCD  partition functions and a 4-point  $q$-deformed correlators.
In particular the analytic continuation  (\ref{highermas}) will be mapped to the  analytic continuation of the momentum of one of the  primaries  to a  higher level degenerate momentum.
Clearly the degeneration mechanism of 5d partition functions, due to the analytic continuation of the masses,
is not limited to the SCQCD, but it extends to more general quiver gauge theories 
which via the  5d/$q$-CFT correspondence are mapped to higher point correlation functions.

\section{5d partition functions as $q$-correlators}

In \cite{Nieri:2013yra} it was proposed that the 5d partition functions discussed in the previous section are captured by a novel class of correlators, dubbed $q$-correlators or  $q$-CFT correlators,
with underling   $q$-deformed Virasoro symmetry (${\cal V}ir_{q,t}$).
\footnote{
The flux-trap realisation of  3d and 5d theories \cite{or}, should allow to gather evidences of the appearance
of  a  $q$-deformed version of Liouville theory  from M5 branes compactifications, along the lines of  \cite{ft}.
}

Deformed Virasoro  and $\mathcal{W}_{N}$  algebras  were introduced in \cite{Shiraishi:1995rp, fr,Feigin:1995sf, Awata:1995zk}, using a correspondence between  singular vectors and multivariable orthogonal symmetric polynomials or using the Wakimoto realization at the critical level. It  was also independently shown  that the deformed Virasoro algebra emerges as a   symmetry in study of  the Andrews-Baxter-Forester (ABF) model \cite{Lukyanov:1995gs, Lukyanov:1996qs}.

Considering expressions (\ref{good2cft}) and (\ref{idcft2}) for partition functions, the match to $q$-correlators
requires to identify 
\be
~~~~~~~~~{ Z}_{\text{1-loop}}(\sigma,\vec m)\, \Leftrightarrow\, {\rm 3-point~function~factors}\,,
\ee
and 
\be
\,{\cal F}={\cal Z}_{\rm cl} ~{\cal Z}_{\rm inst}\, \Leftrightarrow\, {\cal V}ir_{q,t} {\rm~chiral~blocks}\,.
\ee
The connection between ${\cal V}ir_{q,t}$ chiral blocks  and  the 5d Nekrasov instanton function on $\mathbb{R}^4\times S^1$ was discussed already in the context of the 5d AGT correspondence \cite{Awata:2009ur,Awata:2010yy,Schiappa:2009cc,Mironov:2011dk,Yanagida:2010vz}.\footnote{See \cite{Itoyama:2013mca, Aganagic:2013tta, Tan:2013xba, Bao:2013pwa} for recent developments.}
Our proposal includes the  interpretation of   ${\cal Z}_{\rm cl}$ as the $q$-deformation  of the factor
fixed by the conformal Ward identities.\footnote{For a study of $q$-deformed   $SU(1,1)$ Ward identities see \cite{Bernard:1989jq,quasi2}.}
Indeed, it  is easy to check that since
$$
\lim_{e_3 \to \infty}\Gamma_{q,t}(x/e_3)\simeq e^{-\frac{2\pi}{3!} B_{33}(x|\vec e)}\,,
$$
in the undeformed Virasoro limit, that is  $e_2=b_0=1/e_1$ and $e_3 =\frac{2\pi i}{\beta}\to \infty$, the classical term ${\cal Z}_{\rm cl}$, upon identifying  the Coulomb branch parameter  with  the internal momentum in the correlator  as $\alpha=Q_0/2+a$,
reduces to:
\ben
{\cal Z}_{\rm cl}\to  e^{-\frac{2\pi i}{e_1e_2e_3}\frac{a^2}{\tilde g^2}}=z^{-a^2}=z^{\alpha (Q_0-\alpha)} z^{-Q_0^2/2}\,,
\een
which is the part depending on the internal momentum of the conformal factor multiplying chiral  Virasoro conformal blocks.

In  \cite{Nieri:2013yra}  a novel class of non-chiral, modular  invariant  $q$-correlators was defined and proved to capture the  {\it full } partition function of 5d theories on  compact manifolds.
In particular, the two cases $S^5$ and $S^4\times S^1$ are related to $q$-CFT correlators with symmetry given by different tensor products of  $\mathcal{V}ir_{q,t}$ algebra and  different 3-point functions. In details:
\begin{itemize}
\item
{\bf Squashed $\bf{S^5}$}: correlators have symmetry $(\mathcal{V}ir_{q,t})^3$
and 3-point function\footnote{Where $2\alpha_T=\alpha_1+\alpha_2+\alpha_3$.} 
\be
\label{3S}
C_{S}(\alpha_3,\alpha_2,\alpha_1)=\frac{1}{S_3(2 \alpha_T-E)}\prod_{i=1}^3 \frac{S_3(2 \alpha_i)}{S_3(2\alpha_T-2 \alpha_i)}\,.
\ee
This $q$-CFT was dubbed $S$-CFT in \cite{Nieri:2013yra}, since  degenerate correlators reproduce partition functions for   3d theories on  $S^3_b$ and, as reviewed  in the previous section,  in the block factorized expression of the partition function, the blocks are glued by an $S$-pairing \cite{hb}.
\item {\bf{ $\bf{S^4\times S^1}$}}: correlators have symmetry $(\mathcal{V}ir_{q,t})^2$
and 3-point function
\be
\label{3id}
C_{id}(\alpha_3,\alpha_2,\alpha_1)=\frac{1}{\Upsilon^\beta(2 \alpha_T-Q_0)}\prod_{i=1}^3 \frac{\Upsilon^\beta(2 \alpha_i)}{\Upsilon^\beta(2\alpha_T-2 \alpha_i)}\,.
\ee
This $q$-CFT was dubbed $id$-CFT  in \cite{Nieri:2013yra}, since  degenerate correlators are equivalent to 3d partition functions on $S^2\times S^1$, that in the block factorized expression involves an $id$-pairing \cite{hb}.

\end{itemize}
The 3-point functions were derived in \cite{Nieri:2013yra}   by means of the bootstrap approach,  studying the crossing symmetry invariance of  4-point correlators with the insertion of a level-2 degenerate state. These degenerate correlators were in turn  argued to reproduce partition functions of certain 3d theories defined on codimension-2 submanifolds of the 5d space. 
The degenerate correlators were constructed in \cite{Nieri:2013yra} exploiting modular invariance and the fact that they are bounded to satisfy  certain difference equations. However, these correlators can also be obtained as a limit of  non-degenerate correlators, analytically continuing the momenta of the states to degenerate values. This is the $q$-CFT analogue of the degeneration mechanism discussed in the previous section, that permits to obtain 3d partition functions as a limit of 5d partition functions.  We will focus on  this limit later in this section.

In  \cite{Nieri:2013yra} the 4-point non-degenerate  correlator on the 2-sphere was analyzed and  identified with the $\mathcal{N}=1$ 5d $SU(2)$ theory with four fundamental flavors. Below we give another example showing how the 1-point torus correlator captures the $\mathcal{N}=1^*$  $SU(2)$ theory, that is the theory of a vector multiplet coupled to one adjoint hyper of mass $M$.
Since the map between the  5d instanton partition functions and $\mathcal{V}ir_{q,t}$ blocks is established we 
focus only on the 1-loop part of the partition function.

\subsection{An example: Torus with one puncture}
\begin{itemize}

\item {\bf $id$-CFT}:
For this $q$-CFT, the 3-point function is given in formula (\ref{3id}). As in standard 2d CFT, we assume that the  correlator can be decomposed as a product of 3-point functions and  $q$-deformed conformal blocks. In particular, denoting by $\alpha$ the internal state and by $\alpha_1$ the external puncture, the 3-point function contribution to the 1-point torus correlator can be written as 
\ben\label{tid}
 C_{id}(Q_0-\alpha,\alpha_1,\alpha)&=&
\frac{\Upsilon^{\beta}(2\alpha_1)}{\Upsilon^{\beta}(\alpha_1)\Upsilon^{\beta}(\alpha_1)}\frac{\Upsilon^{\beta}(-Q_0+2\alpha)\Upsilon^{\beta}(Q_0-2\alpha)}{\Upsilon^{\beta}(\alpha_1-Q_0+2\alpha)\Upsilon^{\beta}(\alpha_1+Q_0-2\alpha)}\,.
\een
As usual we relate the  internal state $\alpha$ to the gauge theory Coulomb branch parameter and the external state $\alpha_1$  to the mass of the adjoint hyper multiplet. The precise dictionary reads
\ben
\alpha=\frac{Q_0}{2}+a\,, \qquad\qquad  \alpha_1=\frac{Q_0}{2}+iM\,,
\een
and, up to factors independent on the Coulomb branch parameters, the  (\ref{tid}) can be written as 
\ben
 C_{id}(Q_0-\alpha,\alpha_1,\alpha)&\sim&
\frac{\Upsilon^{\beta}(2a)\Upsilon^{\beta}(-2a)}{\Upsilon^{\beta}(\frac{Q_0}{2}+iM+2a)\Upsilon^{\beta}(\frac{Q_0}{2}+iM-2a)}\,.
\een
This is the $S^4\times S^1$ 1-loop contribution for an $SU(2)$ vector coupled to an adjoint hyper, multiplied by the Vandermonde, confirming that the 1-point torus correlator is related to the 5d $\mathcal{N}=1^*$  $SU(2)$ theory.   

\item {\bf $S$-CFT}:
As in the previous example, we name $\alpha$ the internal state and $\alpha_1$ the external state.   The 
3-point function contribution for the 1-punctured torus is given by 
\ben
 C_{S}(E-\alpha,\alpha_1,\alpha)&=&
\frac{S_3(2\alpha_1)}{S_3(\alpha_1)S_3(\alpha_1)}\frac{S_3(-E+2\alpha)S_3(E-2\alpha)}{S_3(\alpha_1-E+2\alpha)S_3(\alpha_1+E-2\alpha)}\,.
\een
Relating $q$-CFT quantities to gauge theory quantities as 
\ben
\alpha=\frac{E}{2}+a\,, \qquad\qquad  \alpha_1=\frac{E}{2}+iM\,,
\een
we obtain
\ben
 C_{S}(E-\alpha,\alpha_1,\alpha)&\sim&
\frac{S_3(2a)S_3(-2a)}{S_3(\frac{E}{2}+iM+2a)S_3(\frac{E}{2}+iM-2a)}\,,
\een
that is the $S^5$ 1-loop contribution of an $SU(2)$ vector and an adjoint hyper of mass $M$, multiplied by the Vandermonde.  Also for the $S$-CFT case, we confirm that the 1-punctured torus correlator is related to the  5d $\mathcal{N}=1^*$  $SU(2)$ theory.
\end{itemize}

 It is worth noting when the adjoint mass is analytically continued to the particular values $M_{}=\frac{i}{2}(Q_0-2b_0)$ or $M_{}=\frac{i}{2}(\omega_1+\omega_2-\omega_3)$ for the $S^4\times S^1$ and $S^5$ theories respectively, the vector and the adjoint almost completely simplify each other leaving a $q$-deformed Vandermonde only. 
 This kind of simplification has already been observed in \cite{Kim:2012ava} and interpreted as global symmetry enhancement, and further studied in \cite{Minahan:2013jwa}.

\subsection{Degeneration of  $q$-correlators}

\begin{figure}[!ht]
\leavevmode
\begin{center}
\includegraphics[width=0.7\textwidth]{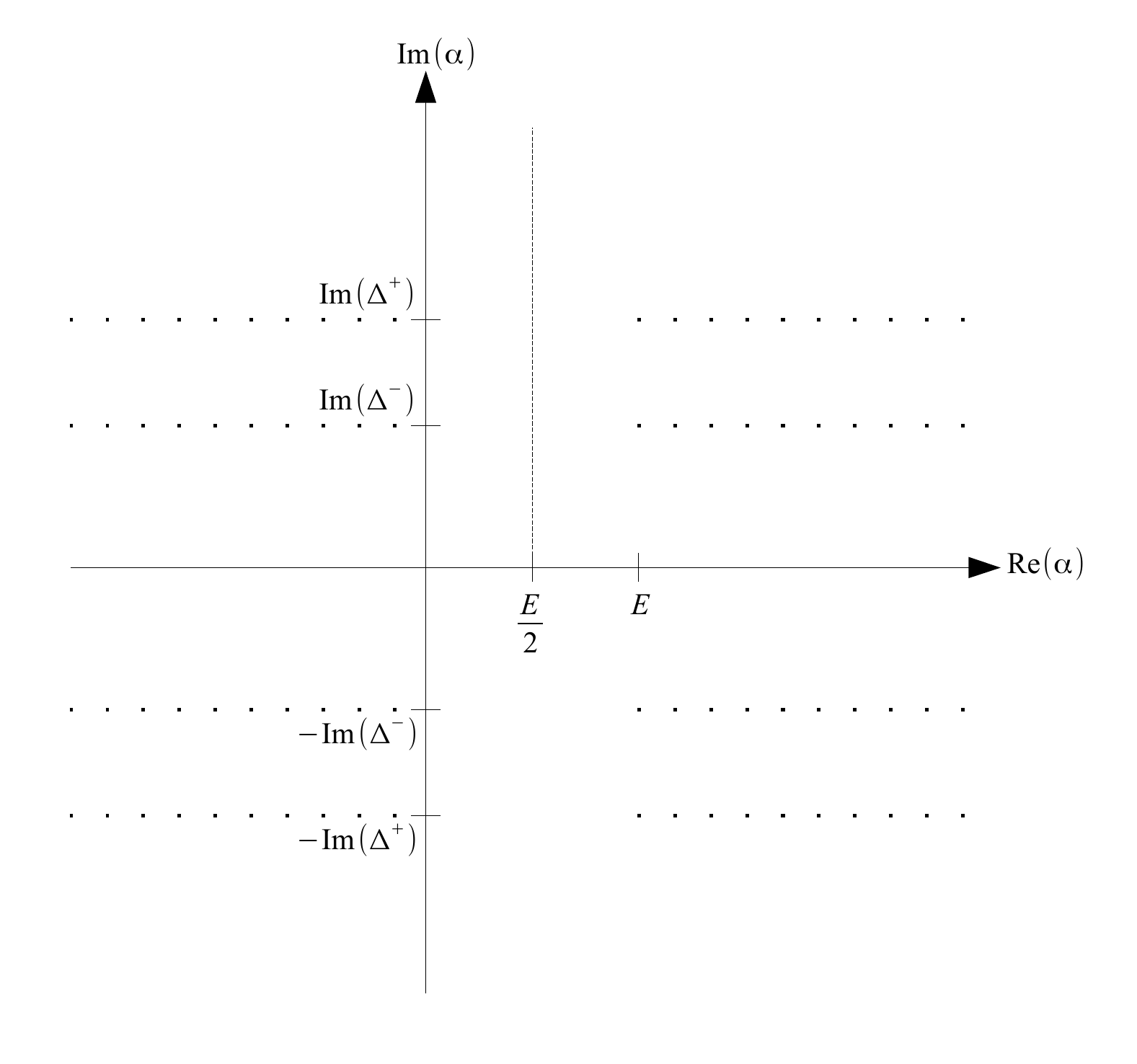}
\end{center}
\caption{Integration path for the fusion of two non-degenerate states.}
\label{pinch1}
\end{figure}

We now study the degeneration of $q$-CFT correlators, that is, we consider the case where the momenta of the states are analytically continued to degenerate values, corresponding to degenerate representations of the $q$-deformed Virasoro algebra. 
In particular, we are interested in determining the set of internal states in the case when one of the external states is degenerate. 
To this end, we analyze the OPE for $q$-CFT  states and study the limit where one of the  ingoing states assumes a degenerate momentum.    We focus here on  the  $S$-CFT, reminding that for this theory, the  
3-point function of non-degenerate primaries is given by
\be\label{tpoint}
C_S(\alpha_2,\alpha_1,\alpha)=\frac{S_3(2\alpha_2)S_3(2\alpha_1)S_3(2\alpha)}{S_3(\alpha_1+\alpha_2+\alpha-E)S_3(-\alpha_1+\alpha_2+\alpha)S_3(\alpha_1-\alpha_2+\alpha)S_3(\alpha_1+\alpha_2-\alpha)}\,.
\ee
In analogy with the  standard CFT case, we can obtain fusion rules between primaries in terms of the 3-point function 
\be\label{fus}
 V_{\alpha_2}(z) V_{\alpha_1}(0)\simeq \int\rd \alpha\;C_S(\alpha_2,\alpha_1,\alpha)[V_{E-\alpha}](z)
\ee
for $z\to 0$.\footnote{ A similar computation for the case of Liouville and $H^+_3$ theory is carefully described in \cite{Jego:2006ta}.} As discussed in appendix \ref{tri},  the triple sine function $S_3$ has an infinite set of zeros distributed  along two semi-infinite lines separated by an interval $E$. This implies that  the 3-point function  (\ref{tpoint}), as a function of the variable $\alpha$, has an infinite number of poles and an infinite number of zeros. In particular, the  poles are distributed along four pairs of semi-infinite lines, each separated by $E$. In details, they are located at
\ben
\alpha&=&\left\{\begin{array}{ll}-\Delta^+ +E-\vec n\cdot\vec \omega;&\qquad -\Delta^+ +2E+\vec n\cdot\vec \omega\\
\phantom{-}\Delta^- -\vec n\cdot\vec \omega;&\qquad \phantom{-} \Delta^- +E+\vec n\cdot\vec \omega\\
-\Delta^- -\vec n\cdot\vec \omega;&\qquad-\Delta^- +E+\vec n\cdot\vec \omega\\
\phantom{-}\Delta^+ -E-\vec n\cdot\vec \omega;&\qquad \phantom{-}\Delta^+ +\vec n\cdot\vec \omega
\end{array}\right.
\een
where we defined $\Delta^\pm=\alpha_1\pm\alpha_2$,  $\vec n\cdot\vec \omega=n_1\omega_1+n_2\omega_2+n_3\omega_3$ and $n_1,n_2,n_3$ are non-negative integers. The zeros are located at 
\be
\alpha=\{-\vec n\cdot\vec \omega/2\,,\quad E/2+\vec n\cdot\vec \omega/2\}\,.
\ee
In the case where $\alpha_1$ and $\alpha_2$ are non-degenerate states it results $\text{Re}(\alpha_1)=\text{Re}(\alpha_2)=E/2$ and the 3-point function (\ref{tpoint})  is analytic in the strip $\text{Re}(\alpha)\in(0,E)$, see Fig. \ref{pinch1}. The integration in formula (\ref{fus}) is performed along the path $\alpha=E/2+i\mathbb{R}^+$ without encountering any pole,  which implies that the OPE of two non-degenerate states produces  a complete set  of non-degenerate states.  This is in agreement with the fact that in the bootstrap decomposition of non-degenerate correlators,  internal channels include the full spectrum of non-degenerate states. Indeed, the internal states result form the fusion of  external non-degenerate states.

\begin{figure}[!ht]
\leavevmode
\begin{center}
\includegraphics[width=0.7\textwidth]{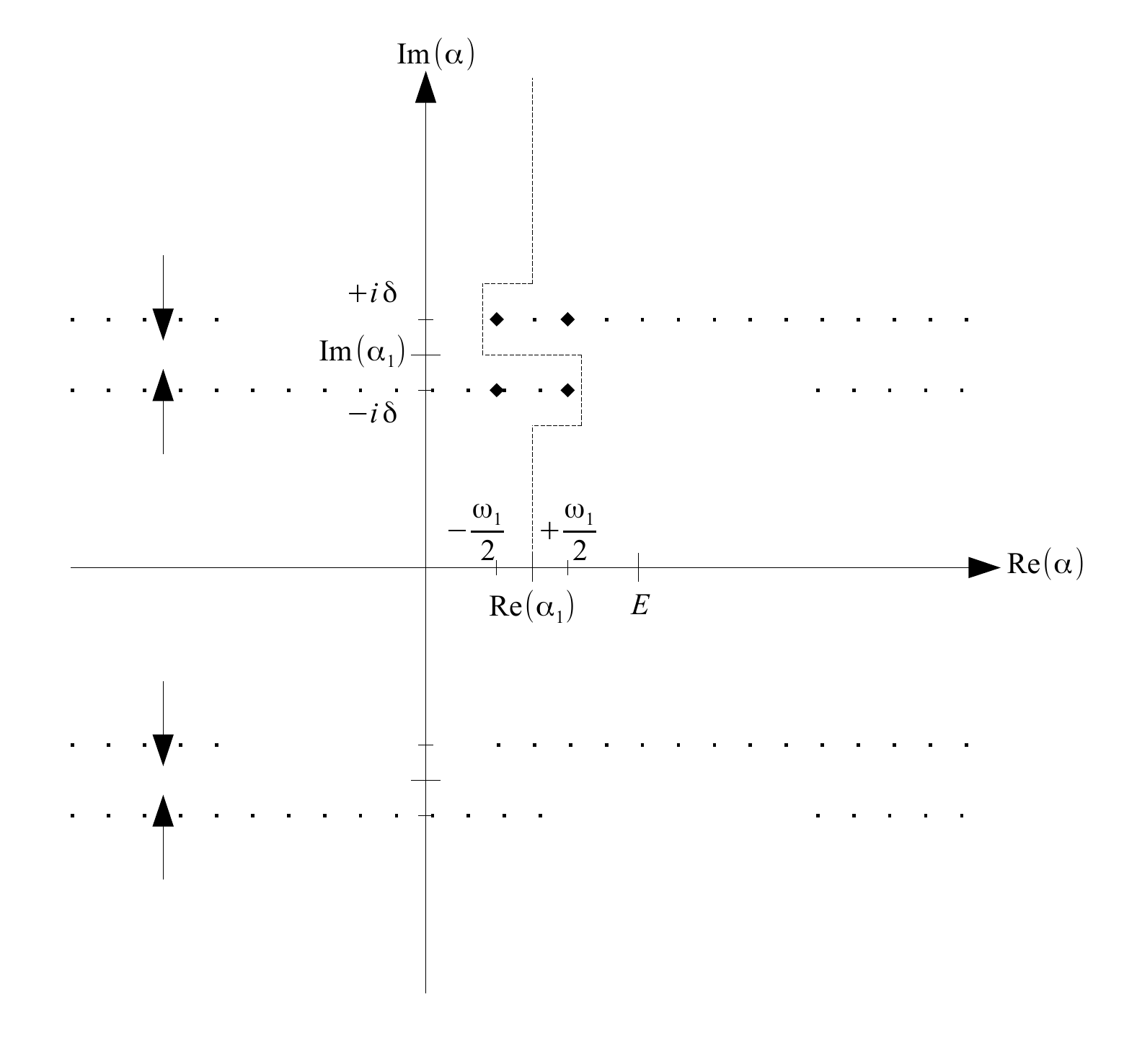}
\end{center}
\caption{Pinching of the integration contour as the 3-point functions are continued to the degenerate values.}
\label{pinch2}
\end{figure}

We now consider the case where one of the states in the OPE,  say $\alpha_2$, is associated to a  degenerate representation, \emph{i.e.}  $\alpha_2=-\vec n\cdot\vec \omega/2$ for a  certain set of non-negative integers $n_1,n_2,n_3$.  This OPE  is computed via  meromorphic analytical continuation as shown in  \cite{Ponsot:2000mt}, that is  we set $\alpha_2=-\vec n\cdot\vec \omega/2+i\delta$ and consider the limit  $\delta\rightarrow 0$. In this limit, due to the factor $S_3(2\alpha_2)$ in the numerator of the 3-point functions, the OPE vanishes on the complex plane, except on the points where the denominator of the 3-point function becomes singular.  As shown in Fig. \ref{pinch2}, the integration path in  (\ref{fus}) is deformed, and the integral receives contribution only from the discrete set of points where the denominator develop double poles, that are located at
\be
\alpha=\alpha_1-\vec s\cdot\vec \omega/2\quad\quad\text{for any}\quad s_k\in  \{-n_k,-n_k+2,\ldots,n_k-2,n_k\}\,   \quad\text{where}\quad k=1,2,3.
\ee
The result is computed by picking the residues at these poles, and shows that the OPE of a non-degenerate state with a degenerate state, include only a finite set of primaries, as in standard CFT.  For instance, in the simplest case where we take $n=(0,0,1)$ (that is, $\alpha_2=-\omega_3/2$),   there are only two contributing poles located at 
\be\label{intsta}
\alpha=\alpha_1\pm\omega_3/2\,.
\ee
This produces the fusion rule
\be
[\alpha]\times [-\frac{\omega_3}{2}]=[\alpha-\frac{\omega_3}{2}]+[\alpha+\frac{\omega_3}{2}]\,,
\ee
which is  analogous to the well-know fusion rule between a level-2 degenerate state and a non-degenerate state in Liouville CFT.

From these  degenerate fusion rules, we can conclude that the internal channel of a degenerate 4-point function includes only a discrete set of states. In particular, in the case where one of the four external states assumes the lowest degenerate momentum $-\omega_3/2$, the internal channel includes only 2 states. We have encountered this result already in  section \ref{secdeg} in the  gauge theory setup. Indeed, as we have already mentioned,  the degeneration of 5d partition functions to  3d partition functions is the gauge theory realization of the degeneration limit of $q$-correlators, where the degeneration of the external state momentum in  the $q$-correlator corresponds to an analytical continuation  of the mass parameters in the gauge theory. In particular, in the gauge theory degeneration limit, the Coulomb branch integral reduces to the sum over the two points given in (\ref{deg1}), that correspond to the momenta of the two internal states given in (\ref{intsta}). 
Considering a generic degenerate state $\alpha_2=-\vec n\cdot \vec \omega/2$, the total number of contributing double poles is $(n_1+1)(n_2+1)(n_3+1)$ which yields the same number of states in the OPE. 
In the  particular  case of the  4-point correlator with a  degenerate insertion  $\alpha_2=-\vec n\cdot \vec \omega/2$,   the $(n_1+1)(n_2+1)(n_3+1)$ internal states  should correspond to the independent solutions of a certain difference operator. This is the $q$-CFT analogue of the higher degeneration of $S^5$ partition functions described in section \ref{sechdeg}.

\section{Reflection coefficients}

As we reviewed in the previous section, the  two families of deformed Virasoro theories which  we have been discussing, the $id$-CFT and $S$-CFT,  were constructed in \cite{Nieri:2013yra}   by means of the  bootstrap method. This approach  does not rely on the Lagrangian formulation of the theory which in the present case is indeed  not available and it is purely   axiomatic since it uses the representation theory of the deformed Virasoro algebra and an ansatz  for the way ${\cal V}ir_{q,t}$ blocks are paired to construct correlators.

However it would still  be useful  to develop a more physical or at  least a geometrical  (rather than purely algebraic) understanding of these theories. For this reason in this section we focus on reflection coefficients. 

In Liouville field theory (LFT) the reflection coefficient  can be   defined in terms of 3-point functions, hence derived  axiomatically from the bootstrap approach. However, it  can also be obtained  from a semiclassical analysis studying  the reflection  from the Liouville wall. In this sense  the reflection coefficient  bridges  between the axiomatical and the semiclassical approach and appears to be an interesting object to study in the
$q$-deformed cases.
Furthermore, as we will see, reflection coefficients constructed from $id$-CFT and $S$-CFT 3-point functions are different, hence they  are sensitive to the way ${\cal V}ir_{q,t}$ blocks are paired to construct correlators.
Therefore we expect that the study of reflection coefficients will help us to
put into perspective  the relation between the two families of $q$-Virasoro systems and LFT.

\subsection{Liouville Field Theory}
In LFT the reflection coefficient is defined as the following ratio of DOZZ 3-point functions {\cite{Zamolodchikov:1995aa}}
\be\label{RL}
R_L(P)=\frac{C(Q_0-\alpha,\alpha_2,\alpha_1)}{C(\alpha,\alpha_2,\alpha_1)}\,, \quad P=i\alpha-iQ_0/2\,,
\ee
where $Q_0=b_0+1/b_0$ and $b_0$ is the Liouville coupling. Using the DOZZ formula for 3-point functions
 \cite{Dorn:1992at,Dorn:1994xn,Zamolodchikov:1995aa}\footnote{We are neglecting a prefactor, which is not important for the present analysis. 
} 
\be
C(\alpha_3,\alpha_2,\alpha_1)=\frac{\Upsilon(2\alpha_3)\Upsilon(2\alpha_2)\Upsilon(2\alpha_1)}
{\Upsilon(\alpha_1+\alpha_2+\alpha_3-Q_0)\Upsilon(-\alpha_1+\alpha_2+\alpha_3)\Upsilon(\alpha_1-\alpha_2+\alpha_3)\Upsilon(\alpha_1+\alpha_2-\alpha_3)}\,,
\ee
one finds:
\be
\label{rexact}
R_{L}(P)=\frac{\Upsilon(-2iP)}{\Upsilon(2iP)}=-\frac{\Gamma(2iPb_0)\Gamma(2iP/b_0)}{\Gamma(-2iPb_0)\Gamma(-2iP/b_0)}\,.
\ee

The reflection coefficient can be also obtained from a semiclassical analysis known an as the mini-superspace approximation \cite{Seiberg}, 
where one ignores oscillator modes and   focuses  on the dynamics of the zero-mode $\phi_0$ of the Liouville field, governed by the Hamiltonian { \cite{Zamolodchikov:1995aa}} 
\be\label{LFTH0}
H_0=-\frac{1}{12}-\frac{1}{2}\frac{\partial^2}{\partial \phi_0^2}+2\pi\mu e^{2b_0\phi_0}.
\ee  
The Schr\"odinger equation for the zero-mode of the Liouville field which scatters from an exponential barrier, since at $\phi_0\to -\infty$ the potential vanishes, has  stationary asymptotic solutions of the form
\be
\psi\sim e^{2iP\phi_0}+R(P)e^{-2iP\phi_0}\,.
\ee
Given that the eigenvalue problem
\be\label{LFTschroedinger}
-\partial^2_{\phi_0}\psi_P+4\pi\mu e^{2b_0\phi_0}\psi_P=4P^2\psi_P
\ee
is solved by the modified Bessel function of the first kind, the reflection amplitude $R(P)$ can be easily extracted and reads
\be\label{RLsemicl}
R(P)\sim\frac{\Gamma(2iP/b_0)}{\Gamma(-2iP/b_0)}.
\ee
This is a semi-classical result valid when $b_0\to 0$, indeed it captures only
 a part of the exact reflection coefficient  (\ref{RL}). Therefore $R_L(P)$ can be considered  the full quantum completion (i.e. including the non-perturbative $1/b_0$ contribution) of the reflection amplitude $R(P)$ computed from the 1st quantized Liouville theory.\\

There is also another (geometric) realization of the above problem, related to harmonic analysis on symmetric spaces. Often, 1d Schr\"odinger problems can be  mapped to free motion of  particles in curved spaces, where a potential-like term  arises once the flat kinetic term is isolated from the Laplace-Beltrami operator. This is indeed the case for the Liouville eigenvalue problem (\ref{LFTschroedinger}) which appears in the study of 
the Laplace-Beltrami operator on the hyperboloid
\be
\label{hyper}
\{(X_0,X_1,X_2,X_3)\in\mathbb{R}^{3,1}\Big|\; X_0^2-X_1^2-X_2^2-X_3^2=1\}\,. 
\ee
Parameterising the hyperboloid (\ref{hyper}) by horospherical coordinates $(x,r,\phi)$ as
\be
X_0=\cosh\frac{x}{2}+\frac{r^2}{2}e^{-x/2};\quad X_1=-\sinh\frac{x}{2}-\frac{r^2}{2}e^{-x/2};\quad X_2=re^{-x/2}\cos\phi;\quad X_3=re^{-x/2}\sin\phi\,,
\ee
the Laplace-Beltrami operator defined by
$$
\nabla^2=\frac{1}{\sqrt{{\rm det}~ g}}\partial_i \sqrt{{\rm det}~ g}~g^{ij}\partial_j\,,\qquad i=x,r,\phi\,,
$$
with $g_{ij}=\partial_i \vec{X}\cdot \partial_j \vec{X}$ the induced metric,
yields:
\be
\nabla^2\Phi=\left(4\partial_x^2-4\partial_x+e^{x}\partial^2_r+\frac{e^{x}}{r}\partial_r+\frac{e^{x}}{r^2}\partial^2_\phi\right)\Phi\,.
\ee
Upon the reduction 
$$\Phi(x,r,\phi)=\rho(r)e^{x/2}\psi(x)\,,\qquad \rho''+\rho'/r=-4k\rho\,,$$
the Laplace-Beltrami operator becomes:
\be
\nabla^2\Phi=\rho(r)e^{x/2}\left(4\partial^2_x\psi(x)-4ke^{x}\psi(x)-\psi(x)\right)\,,
\ee
corresponding indeed to the Hamiltonian $H_0$ given in (\ref{LFTH0}).  The free motion $-\nabla^2\Phi=\lambda^2\Phi$ is in turn translated into the Liouville eigenvalue equation  (\ref{LFTschroedinger})
\be\label{LBLobachevsky}
-\partial^2_x\psi_\lambda(x)+k e^{x}\psi_\lambda(x)=\lambda^2\psi_\lambda(x)\,,
\ee
with asymptotic solutions of the form
\be
\psi_\lambda(x)\sim c(\lambda)e^{i\lambda x}+c(-\lambda)e^{-i\lambda x}\quad \text{ as } \quad x\to -\infty
\,.\ee
The coefficients $c(\pm \lambda)$ are known in this context as Harish-Chandra $c$-functions, which for the hyperboloid  (\ref{hyper}) are given by:\footnote{We neglect inessential factors.}
\be
c(\lambda)=\frac{1}{\Gamma(1+2i \lambda)}\,,
\ee
yielding
\be
\label{ss}
\frac{c(-\lambda)}{c(\lambda)}=-\frac{\Gamma(2i\lambda)}{\Gamma(-2i\lambda)}\,,
\ee
which,  with the identification $P/b_0=\lambda$, corresponds  to the semiclassical result (\ref{RLsemicl}).\\

The hyperboloid  (\ref{hyper})  is isomorphic to the  Lobachevsky space $SO_0(1,3)/SO(3) \simeq SL(2,\mathbb{C})/SU(2)$.  In the study of group manifolds the Laplace-Beltrami operator  appears from the analysis of the quadratic Casimirs. In particular, horospheric coordinates can be introduced for any hyperbolic symmetric space, reflecting the Iwasawa decomposition of the group manifold. The plane wave asymptotic behaviour of the eigenfunctions of the reduced Laplace-Beltrami operator is then governed by the  Harish-Chandra $c$-functions as in the case of the hyperboloid.
In fact, $c$-functions of any classical symmetric space can be expressed as a product of Gamma functions ({\it Gindikin-Karpelevich formula}, see for instance \cite{Helgason}, cf. \cite{HelgasonI}). In horospherical coordinates ({\it e.g.} \cite{GerasimovKharchevMarshakovMironovMorozovOlshanetsky}), one has
\be
\label{prodoto}
c(\lambda)=\prod_{\alpha\in \Delta^+}\frac{1}{\Gamma(l+\lambda\cdot\alpha)}\,,
\ee
where $\lambda$ is a spectral parameter for the eigenvalues of the Laplacian, $\Delta^+$ denotes the positive roots of the Lie algebra of the group modelling the symmetric space, and $l=1$ for finite dimensional Lie algebras or $l=1/2$ for the affine version thereof. The reflection coefficient of the associated quantum mechanical system can be computed as the ratio $c(-P)/c(P)$, where the momentum $P$ is parameterised by $\lambda\cdot \alpha$.\\

As we already observed, the semiclassical reflection coefficient $R(P)$ (\ref{ss}) or (\ref{RLsemicl}) reproduces only part of the  exact LFT reflection coefficient $R_L(P)$ (\ref{rexact}).
There is however a group theoretic  method \cite{GerasimovKharchevMarshakovMironovMorozovOlshanetsky} to obtain the non-perturbative or 2nd quantized completion of the LFT reflection coefficient. To this end, one considers the extension of the root system, which  in the  $\mathfrak{sl}(2)$ case contains   only one positive root $\alpha_1$,  by the affine root $\alpha_0$ so that  upon affinization $\mathfrak{sl}(2)\rightarrow \widehat{\mathfrak{sl}}(2)$ the positive roots become:
\be\label{LFTaff}
\alpha_0+n\delta, \quad n\delta, \quad \alpha_1+n\delta, \quad \delta\equiv\alpha_0+\alpha_1, \quad n\in \mathbb{Z}^+_0\,.
\ee
The Harish-Chandra function is then given by the product  (\ref{prodoto}) over  the extended root system and,
choosing the parameterization
\be\label{LFTaffpar}
\lambda\cdot\delta=\tau, \quad \lambda\cdot\alpha_1=2iP/b_0-1/2, \quad \lambda\cdot\alpha_0=\tau-2iP_0/b_0+1/2\,,
\ee
becomes
\ben
\label{cg2}
c(P)^{-1}&\equiv& \prod_{n\geq 0}\Gamma(1/2 + \lambda \cdot(\alpha_1 + n \delta))\prod_{n\geq 0}\Gamma(1/2 + \lambda \cdot(\alpha_0 + n \delta))\prod_{n\geq 1}\Gamma(1/2 + \lambda \cdot n \delta) \nonumber\\
&=&\Gamma(2iP/b_0)\prod_{n\geq 1}\Gamma(2iP/b_0+n\tau)\Gamma(1-2iP/b_0+n\tau)\Gamma(1/2+n\tau)\,.
\een
It is easy to check that the  $n=0$ factor reproduces the non-affinized/semiclassical  result while the ratio
of $c$-functions
\ben
\frac{c(-P)}{c(P)}&=&\frac{\Gamma(2iP/b_0)}{\Gamma(-2iP/b_0)}\prod_{n\geq 1}\frac{\Gamma(2iP/b_0+n\tau)\Gamma(1-2iP/b_0+n\tau)}{\Gamma(-2iP/b_0+n\tau)\Gamma(1+2iP/b_0+n\tau)}\nn\\
&\sim&-\frac{\Gamma(2iP/b_0)}{\Gamma(-2iP/b_0)}\frac{\Gamma(2iP/b_0\tau)}{\Gamma(-2iP/b_0\tau)}\,,
\een
upon identifying  $\tau=1/b_0^2$, reproduces  the exact LFT reflection coefficient $R_L$ (\ref{rexact}).\\

The above discussion suggests to  regard the affinization as a  prescription  for an effective 2nd quantization.
As we are about to see,  this prescription works remarkably well for our $q$-Virasoro systems.


\subsection{$id$-CFT}\label{refid}

We begin by computing  the $id$-CFT reflection coefficient in terms of the $id$-pairing 3-point functions given in (\ref{3id}):
\be
R_{id}(P)=\frac{C_{id}(Q_0-\alpha,\alpha_2,\alpha_1)}{C_{id}(\alpha,\alpha_2,\alpha_1)}=\frac{\Upsilon^\beta(-2iP)}{\Upsilon^\beta(2iP)}\,, \qquad P=i\alpha-iQ_0/2\,,
\ee
which can be rewritten in terms of the $q$-deformed  $\Gamma$ function:
\be
\Gamma_q(x):=\frac{(q;q)}{(q^x;q)}(1-q)^{1-x}\,,
\ee
as
\be
\label{qLiouville}
R_{id}(P)\sim\frac{(1-q)^{4iPb_0}}{(1-t)^{-4iP/b_0}}\frac{1-q^{2iPb_0}}{1-t^{-2iP/b_0}}\times\frac{\Gamma_q\left(2iPb_0\right)\Gamma_t\left(2iP/b_0\right)}
{\Gamma_q\left(-2iPb_0\right)\Gamma_t\left(-2iP/b_0\right)}\,,
\ee
with $q=e^{\beta/b_0}$ and $t=e^{\beta b_0}$.

The  above expression suggests that in this case  the 1st quantized reflection coefficient
should be captured by a $c$-function expressed in terms of products over the $\mathfrak{sl}(2)$ root system  of $q$-deformed (with deformation parameter $t$) Gamma functions
\be\label{c-qLFT}
c(\lambda)=\frac{1}{\Gamma_t(l+\lambda\cdot\alpha)}\,,
\ee
while  the affinization prescription  should allow to recover the non-perturbative reflection coefficient (\ref{qLiouville}). Indeed proceeding as in  the  LFT case (\ref{LFTaff}), (\ref{LFTaffpar}), we obtain
\be
c(P)^{-1}= \prod_{n\geq 0}\Gamma_t(2iP/b_0+n\tau)\Gamma_t(1-2iP/b_0+(n+1)\tau)\prod_{n\geq 1}\Gamma_t(1/2+n\tau)\,,
\ee
which, after dropping $P$-independent factors and defining $q=t^\tau$, becomes
\be
c(P)\sim (t^{2iP/b_0};q,t)(qt^{1-2iP/b_0};q,t)=\frac{(t^{2iP/b_0};q,t)(t^{1-2iP/b_0};q,t)}{(t^{1-2iP/b_0};t)}\,.
\ee
It is  immediate to verify that the ratio of $c$-functions
\ben
\frac{c(-P)}{c(P)}
&=&\frac{(1-q)^{4iPb_0}}{(1-t)^{-4iP/b_0}}\frac{1-q^{2iPb_0}}{1-t^{-2iP/b_0}}\frac{\Gamma_q\left(2iPb_0\right)\Gamma_t\left(2iP/b_0\right)}{\Gamma_q\left(-2iPb_0\right)\Gamma_t\left(-2iP/b_0\right)},
\een
reproduces $R_{id}$ once we set $\tau=1/b_0^{2}$. 
Since $\Gamma_q\to\Gamma$ as $q\to1$, we can also check that, as expected, in the   $\beta\to 0$ limit
the $id$-CFT reflection coefficient $R_{id}$ reduces  to   the LFT one $R_L$.\\

The $c$-function (\ref{c-qLFT}), suggested by the naive $q$-deformation of the Liouville case,
actually appears in \cite{OlshanetskyRogov}  (see also \cite{FreundZabrodinI,FreundZabrodinII,FreundZabrodinIV})
) as the genuine $c$-function in the quantum Lobachevsky space\footnote{The identification follows from $q^{-2}\to t$, $2i\theta\to \lambda\cdot\alpha$.}. In that case the relevant quantum group is $U_q(\mathfrak{sl}(2))$, and the quadratic Casimir can be studied introducing horospherical coordinates in the quantum space. 
The eigenvalue problem and asymptotic analysis then  leads exactly to (\ref{c-qLFT}). In particular the eigenvalue problem from which the $c$-function is derived is a discretized version of the Schr\"odinger-Liouville equation (\ref{LBLobachevsky}).\footnote{This is also very similar to the expression for the Laplacian on the complex $q$-plane that can be found in \cite{Ubriaco}, formula (22). In notations of \cite{Ubriaco} and upon the change of variables $z = log x$, one obtains 
$$
h_0 f(z) = - \frac{1}{z^2 \big(q - \frac{1}{q}\big)} \, \Big[\frac{1}{q} f(qz) + q \, f\big(\frac{z}{q}\big) - \big(q + \frac{1}{q}\big) \, f(z)\Big].
$$}

{ Finally, let us explain why we can expect the affinization procedure to work also in the case of a quantum affine algebra $U_q(\widehat{\alg{g}})$, with $\alg{g}$ a simple Lie algebra. In fact, once we assume a specific form of the $c$-function for the non-affine part $\alg{g}$, we can rely on the fact that the root structure is the same for $U_q(\widehat{\alg{g}})$ as it is for the undeformed affine case $\widehat{\alg{g}}$ \cite{LevendorskiiSoibelmanStukopin}. This is made particularly evident in what is known as {\it Drinfeld's second realization} \cite{Drinfeld} of quantum affine algebras. In this realization, to each of the infinite roots  (\ref{LFTaff}) is associated a particular generator labelled by the integer level $n$. Roots are divided into positive and negative, and naturally ordered according to increasing or decreasing $n$. The specification of the algebra is then completed by assigning a set of relations between the generators at the various levels.
}

\subsection{$S$-CFT}\label{Gamma2}
The  reflection coefficient constructed in terms of the  $S$-CFT 3-point functions 
(\ref{3S}) is given by:
\be
R_S(P)=\frac{C_{S}(E-\alpha,\alpha_2,\alpha_1)}{C_{S}(\alpha,\alpha_2,\alpha_1)}=\frac{S_3(-2iP|\vec \omega)}{S_3(2iP|\vec \omega)}\,,
\quad P=i\alpha-iE/2\,.
\label{rs}
\ee
In this case if we take a  $c$-function given in terms of  Barnes double Gamma functions 
\be
c(P)=\frac{1}{\Gamma_2(2iP/\kappa|e_1,e_2)}\,, \quad e_1+e_2=1\,,
\ee
while keeping the $\mathfrak{sl}(2)$ root system, and apply the affinization prescription we obtain
\ben
c(P)^{-1}&=&\prod_{n\geq 0}\Gamma_2(2iP/\kappa+n\tau|e_1,e_2)\Gamma_2(1-2iP/\kappa+(n+1)\tau|e_1,e_2)\prod_{n\geq 1}\Gamma_2(1/2+n\tau|e_1,e_2)\nn\\
&&\sim\Gamma_3(2iP/\kappa|e_1,e_2,\tau)\Gamma_3(e_1+e_2+\tau-2iP/\kappa|e_1,e_2,\tau)=S_3(2iP/\kappa|e_1,e_2,\tau)^{-1}\,.\nn\\
\een
Finally, using $S_3(\kappa X|\kappa \vec \omega)=S_3(X|\vec \omega)$, we get
\be
c(P)=S_3(2iP|\vec \omega)\,, \quad \vec\omega =\kappa(e_1,e_2,\tau)\,,
\ee
from which it follows that  the ratio $c(-P)/c(P)$ reproduces the exact reflection  coefficient (\ref{rs}).

Notice that the Barnes $\Gamma_2$ function appears already in the affinized
version of the Liouville $c$-function given in eq. (\ref{cg2})  which indeed can be rewritten as
\be
c(P)^{-1}\sim \Gamma_2(2iP|b_0,b_0^{-1})\Gamma_2(Q_0-2iP|b_0,b_0^{-1})\,,
\ee
so in a sense we may regard $R_S$ as arising from a 2nd affinization, or multi-loop algebra of $\mathfrak{sl}(2)$.
Even if  we are not aware of an explicit way to construct a space whose $c$-function is given by $\Gamma_2$ functions, it has been  strongly motivated for example in \cite{FreundZabrodinII} that such generalised $c$-functions  should naturally be associated to  such a construction, and to general families of integrable systems, whose S-matrix building blocks are indeed $\Gamma_n$ functions or deformations thereof (as we saw in the $id$-CFT) (see also {\it e.g.} \cite{Ruijsenaars}).

Before ending this section, let us observe that the reflection coefficients $R_{id}$ and $R_S$ satisfy the unitary condition $R_{id,S}(P)R_{id,S}(-P)=1$ by construction, and have zeros and poles determined by the factors $\Gamma_q$ and $\Gamma_3$ respectively. Moreover, we saw the Lie algebra $\mathfrak{sl}(2)$ played a prominent role in the study of the reflection coefficients through $c$-functions. As we are going to show in the next section, all these elements are of fundamental importance in integrable systems, and it is therefore natural to ask which known models feature the same structures we have just seen.

\section{S-matrices}
In the previous sections we have shown how it is possible to reproduce the reflection coefficients via an affinization procedure starting from putative Harish-Chandra $c$-functions {\cite{OlshanetskyPerelomov,FreundZabrodinIV} (see also \cite{FaddeevH}, comments following formula (217))}. In this section we will connect these coefficients to known scattering matrices of integrable spin-chains and (related) integrable quantum field theories. Typically, the S-matrices are built by taking the ratio of the two {\it Jost functions} $J(u)$ with opposite arguments, as they appear in the plane-wave asymptotics of the scattering wave function:
\begin{eqnarray}
\label{definSmat}
\psi(x) \sim J(- u) e^{i p x} + J(u) e^{- i p x}, \qquad x \to - \infty, \qquad S(u) = \frac{J(u)}{J(-u)},
\end{eqnarray}
with $p$ the momentum of the particle and $u$ the corresponding rapidity (for massive relativistic particles, $E=m \, \mbox{cosh} u$ and $p = m \, \mbox{sinh} u$).

We will find a relationship between known S-matrices/Jost functions computed in the literature and the $c$-functions we have been using, {\it before} the affinization takes place. The affinization 
{\begin{eqnarray}
\alg{sl}(2) \, \longrightarrow \, \widehat{\alg{sl}}(2)
\end{eqnarray}}
produces then the final expression for the reflection coefficients.

\smallskip

The starting point will be the S-matrix for two excitations of the XYZ spin-chain \cite{FaddeevTakhtadjan}. The Hamiltonian for the XYZ chain\footnote{The literature on this topic is extremely vast. We will mainly follow \cite{CauxKonnoSorrellWeston} for the purposes of this section. For recent work on the XYZ chain, see for instance \cite{ErcolessiEvangelistiFranchiniRavanini}.} reads
\begin{eqnarray}
\label{HXYZ}
H = - \frac{1}{4} \sum_{n=1}^N \, \big(J_x \, \sigma_n^x \, \sigma_{n+1}^x + J_y \, \sigma_n^y \, \sigma_{n+1}^y + J_z \, \sigma_n^z \, \sigma_{n+1}^z\big),
\end{eqnarray}
where the sum is over all the sites of a chain of $N$ sites, and the spin at each site is $\frac{1}{2}$. We consider definite $z$-component of the spin, with the operators $\sigma_n^i$ being the Pauli matrices at site $n$. 

The connection with the integrable 8-vertex model \cite{BaxterP} is obtained by imposing the following parameterization in terms of Jacobi elliptic functions:
\begin{eqnarray}
\label{Bparam}
&&J_x = J \, \big[\mbox{cn}^2(\lambda, k') + k \, \mbox{sn}^2(\lambda, k')\big], \qquad J_y = J \, \big[\mbox{cn}^2(\lambda, k') - k \, \mbox{sn}^2(\lambda, k')\big] \nonumber \\
&&J_z = - J \, \mbox{dn}(\lambda, k'), \qquad \qquad k' \equiv \sqrt{1 - k^2},
\end{eqnarray}
where we will assume $J >0$, and the {\it modulus} $k$, the {\it complementary modulus} $k'$ and the {\it argument} $\lambda$ to be {\it a priori} complex\footnote{The properties of the elliptic functions that we will need here can be found in appendix A of \cite{CauxKonnoSorrellWeston} or in \cite{GradshteynRyzhik}.}. In this fashion, the Hamiltonian (\ref{HXYZ}) can be directly obtained (apart from an overall factor and a constant shift) from the transfer matrix of the 8-vertex model by taking its logarithmic derivative \cite{Baxter}, in the spirit of the quantum inverse scattering method (cf. also \cite{Faddeev} and appendix \ref{BaxtOp}).

One starts focusing on a particular region of the parameter space in the Hamiltonian (\ref{HXYZ}), corresponding to the so-called {\it principal regime}
\begin{eqnarray}
\label{principal}
|J_y| \leq J_x \leq - J_z.
\end{eqnarray}
Any other region of the parameter space can be reached starting from (\ref{principal}) and performing suitable transformations \cite{Baxter}. In this regime, let us recast the Hamiltonian in the following form\footnote{Upon using formulas (15.7.3a/b/c) in \cite{BaxterBook}, and the parity properties of the Jacobi elliptic functions, one can show that the parameterization (10.4.17), (10.15.1b), relevant for the treatment of the XYZ chain in \cite{BaxterBook}, coincides with (\ref{Bparam}). The modulus $k$ in \cite{BaxterBook} is the same as $k$ here.}:
\begin{eqnarray}
\label{HXYZZ}
&&H = - \frac{J_x}{4} \sum_{n=1}^N \, \big(\sigma_n^x \, \sigma_{n+1}^x + \Gamma \, \sigma_n^y \, \sigma_{n+1}^y - \Delta \, \sigma_n^z \, \sigma_{n+1}^z\big), \nonumber\\
&&\Gamma \equiv \frac{J_y}{J_x}, \qquad \Delta = - \frac{J_z}{J_x}, \qquad J_x \geq 0, \qquad \Delta \geq 1.
\end{eqnarray}
One can see from $J_x \geq 0$ and $\Delta \geq 1$ that the chain is in an {\it antiferromagnetic} region, where the alignment of spins along the $z$-axis is energetically disfavoured. The ground state is a Dirac sea of filled levels over the false vacuum (which is the ferromagnetic one with all spins aligned), and the excitations are holes in the sea. These excitations scatter\footnote{The scattering matrices for excitation-doublets in the {\it disordered} regime in particular have been derived in \cite{FioravantiRossiI,FioravantiRossiII}, where interesting connections with the deformed Virasoro algebra are also pointed out.} with a well defined S-matrix, which can in principle be obtained using the method of Korepin \cite{Korepin}. In \cite{FreundZabrodinIV}, the corresponding Jost function is written in terms of parameters $\gamma$ and $\tau$ as an infinite product of $\Gamma_q$ functions with shifted arguments, specifically
\begin{eqnarray}
\label{XYZ}
J(u)\, = \, \prod_{m=0}^\infty \, \frac{\Gamma_q (i u + r m)\Gamma_q (i u + r m + r + 1)}{\Gamma_q (i u + r m + \frac{1}{2})\Gamma_q (i u + r m + r + \frac{1}{2})},
\end{eqnarray}
where 
\begin{eqnarray}
\label{q}
q \, = \, e^{- 4 \gamma}, \qquad r \, = \, - \frac{i \pi \tau}{2 \gamma}, 
\end{eqnarray}
$u$ being a spectral parameter equal to the difference of the incoming particle rapidities $u = u_1 - u_2$, and we should set\footnote{See appendix \ref{parame} for the relationships amongst the various parameters used here, and with the parameters traditionally used in the literature.} $\tau = \frac{i}{2} \frac{K'}{K}$, where $K$ and $K'$ (called, respectively, $I$ and $I'$ in \cite{BaxterBook}) are the complete elliptic integrals of the first kind
$$
K = \int_0^{\frac{\pi}{2}} \frac{dt}{\sqrt{1 - k^2 \sin^2 t}}, \qquad K' = \int_0^{\frac{\pi}{2}} \frac{dt}{\sqrt{1 - k'^2 \sin^2 t}}, \qquad k'^{\, 2} = 1 - k^2.
$$ 

Our two reflection coefficients corresponding, respectively, to the $id$- and to the $S$-pairing, as obtained in the previous sections, can be related to different limits of the above Jost function before the affinization procedure.

\begin{itemize}

\medskip

\item {\bf Limit 1. Reproducing the special functions of the $id$-pairing}

Following \cite{CauxKonnoSorrellWeston}, by sending $k \to 0$ with $\lambda$ fixed and real in (\ref{Bparam}) one obtains the following limit:
\begin{eqnarray}
J_x \to J \, \mbox{sech}^2\lambda \geq 0, \qquad J_y \to J \, \mbox{sech}^2\lambda, \qquad J_z \to - J \, \mbox{sech}\lambda,
\end{eqnarray}
therefore the Hamiltonian reduces to
\begin{eqnarray}
\label{HXXZ}
H = - \frac{J}{4} \mbox{sech}^2\lambda \, \sum_{n=1}^N \, \big(\sigma_n^x \, \sigma_{n+1}^x + \, \sigma_n^y \, \sigma_{n+1}^y - \mbox{cosh}\lambda \, \sigma_n^z \, \sigma_{n+1}^z\big).
\end{eqnarray}
One can see that $\Gamma \to 1$ and $\Delta \to \mbox{cosh}\lambda \geq 1$ for real $\lambda$. This means that the limiting chain is an XXZ spin-chain in its {\it antiferromagnetic} regime. Moreover, the regime $\Delta \geq 1$ is {\it massive}, meaning that the spectrum of excitations (holes in the Dirac sea) has a mass gap.
 
The limit $k \to 0$ corresponds to sending $\tau \to i \infty$, since $K' \to + \infty$ and $K \to \frac{\pi}{2}$. If one performs this limit in the expression (\ref{XYZ}) \cite{FreundZabrodinIV}, only part of the $k=0$ term survives the limit and one obtains \cite{FreundZabrodinI,FreundZabrodin} (up to an overall factor) the Jost function for the scattering of a kink and an anti-kink in the {\it antiferromagnetic} spin-$\frac{1}{2}$ XXZ spin-chain, in terms of a ratio of two $\Gamma_q$ functions 
\begin{eqnarray}
\label{XXZ}
J(u) \, \to \, \frac{\Gamma_q (i u)}{\Gamma_q (i u + \frac{1}{2})}.
\end{eqnarray}
The same scattering factor is also obtained (in a slightly different parameterization) in \cite{DaviesFodaJimboMiwaNakayashiki}, as a multiplier of the R-matrix for a $U_q(\widehat{\mathfrak{sl}}(2))$ doublet of excitations of the antiferromagnetic massive XXZ spin-chain.

The same structure features in our $id$-pairing reflection coefficient. As we recalled in section \ref{refid}, the authors of \cite{OlshanetskyRogov} derive the Harish-Chandra function (\ref{XXZ}) by studying the quantum Lobachevsky space  and already point out the connection to the XXZ quantum spin-chain via (\ref{XXZ}). 

\medskip

\item {\bf Limit 2. Reproducing the special functions of the $S$-pairing}

The second limit considered in \cite{CauxKonnoSorrellWeston} is composed of two operations. Firstly, one performs a transformation that maps the parameters of the Hamiltonian (\ref{HXYZ}) as follows:
\begin{eqnarray}
J_x \to J_x' = - J_z, \qquad J_y \to J_y' = J_x, \qquad J_z \to J_z' = - J_y
\end{eqnarray} 
The above transformation maps the principal regime of the Hamiltonian (\ref{HXYZ}) to its {\it disordered regime}:
$$
|J_z'| \leq J_y' \leq J_x'.
$$
Secondly, by sending now $k \to 1$ with $\lambda$ fixed in (\ref{Bparam}) one obtains the following limit\footnote{The Jacobi functions of modulus $k' = \sqrt{1 - k^2}$ reduce according to $\mbox{sn}(\theta,0) = \sin \theta$, $\mbox{cn}(\theta,0) = \cos \theta$, $\mbox{dn}(\theta,0) = 1$ when the modulus $k'$ goes to $0$.}:
\begin{eqnarray}
J_x' = - J_z \to J, \qquad J_y' = J_x \to J, \qquad J_z' = - J_y \to - J \cos 2 \lambda,
\end{eqnarray}
therefore the Hamiltonian reduces to
\begin{eqnarray}
\label{HXXZe}
H = - \frac{J}{4} \, \sum_{n=1}^N \, \big(\sigma_n^x \, \sigma_{n+1}^x + \, \sigma_n^y \, \sigma_{n+1}^y - \cos 2 \lambda \, \, \sigma_n^z \, \sigma_{n+1}^z\big).
\end{eqnarray}
One can see that $\Gamma \to 1$ and $|\Delta| \to |\cos 2 \lambda| \geq 1$ for real $\lambda$. This means that the limiting chain is an XXZ spin-chain in its {\it disordered} regime. The regime $|\Delta| \leq 1$ is {\it massless}, meaning that the spectrum of excitations (holes in the Dirac sea), called ``spinons'' in this case (with spin equal to $\frac{1}{2}$), has no mass gap. For $\Delta =0$ the excitations are described by a theory of free fermions. 

The limit $k \to 1$ corresponds to sending $\tau \to 0$, since $K \to + \infty$ and $K' \to \frac{\pi}{2}$ in this case. If one performs this limit in the expression (\ref{XYZ}), as pointed out in \cite{FreundZabrodinIV}, namely one first sends $q \to 1$ with $r$ fixed\footnote{See appendix \ref{parame} for the relationships amongst the various parameters, and with those used in the literature.}, one obtains an infinite product of ordinary $\Gamma$ functions:
\begin{eqnarray}
\label{gammante}
J(u)\, \to \, \prod_{m=0}^\infty \, \frac{\Gamma (i u + r m)\Gamma (i u + r m + r + 1)}{\Gamma (i u + r m + \frac{1}{2})\Gamma (i u + r m + r + \frac{1}{2})}.
\end{eqnarray}
The corresponding S-matrix is calculated according to (\ref{definSmat}). One can rewrite the resulting infinite product of $\Gamma$ functions as
\begin{eqnarray}
\label{sinoGordo}
S &=& \frac{J(u)}{J(-u)}\, \to \nonumber\\
&\to& \prod_{n=0}^\infty \frac{\Gamma \Big( \frac{1}{2} + n \, \Sigma - \frac{v}{2}\Big)}{\Gamma \Big( \frac{1}{2} + n \, \Sigma + \frac{v}{2}\Big)} \frac{\Gamma \Big( 1 + n \, \Sigma + \frac{v}{2}\Big)}{\Gamma \Big( 1 + n \, \Sigma - \frac{v}{2}\Big)} \frac{\Gamma \Big((n+1) \, \Sigma + \frac{v}{2}\Big)}{\Gamma \Big((n+1) \, \Sigma - \frac{v}{2}\Big)} \frac{\Gamma \Big( \frac{1}{2} + (n+1) \, \Sigma - \frac{v}{2}\Big)}{\Gamma \Big( \frac{1}{2} + (n+1) \, \Sigma + \frac{v}{2}\Big)}\nonumber\\
&\sim &\frac{\Gamma_2 \Big(\frac{1-v}{2} |1,\Sigma\Big)\Gamma_2 \Big(\frac{1+v}{2} |1,\Sigma\Big)
\Gamma_2 \Big(\frac{v}{2}+\Sigma |1,\Sigma\Big)   \Gamma_2 \Big(\frac{1-v}{2}+\Sigma |1,\Sigma\Big)   }{  \Gamma_2 \Big(\frac{1+v}{2} |1,\Sigma\Big) 
 \Gamma_2 \Big(\frac{1-v}{2} |1,\Sigma\Big)   \Gamma_2 \Big(\Sigma-\frac{v}{2} |1,\Sigma\Big)   \Gamma_2 \Big(\frac{1+v}{2}+\Sigma |1,\Sigma\Big)  \,
 },
\end{eqnarray}
where $\Sigma = r$ and $v = 2 i u$. The $\Gamma_2$ function is the same function we saw playing a r$\hat{\mbox{o}}$le in the $S$-pairing calculation, cf. section \ref{Gamma2}. 
In terms of Jost functions, we have 
\begin{eqnarray}
\label{splitto}
J(u) \to \frac{\Gamma_2 \Big(\frac{1+v}{2} |1,\Sigma\Big)
\Gamma_2 \Big(\frac{v}{2}+\Sigma |1,\Sigma\Big)   }{  \Gamma_2 \Big(\frac{1+v}{2} |1,\Sigma\Big) 
    \Gamma_2 \Big(\frac{1+v}{2}+\Sigma |1,\Sigma\Big)},
\end{eqnarray}
which is the analogue of (\ref{XXZ}).\footnote{We observe that, by defining the combination $
\varphi(v) = \frac{S_2 (\frac{1}{2} + \frac{v}{2}|1,\Sigma)}{S_2 (\frac{v}{2}|1,\Sigma)}$,
we can also rewrite (\ref{sinoGordo}) as $S = \frac{\varphi(v)}{\varphi(-v)}$. A system whose $c$-function is given by $S_2$ has been considered in \cite{KharchevLebedevSemenovTianShansky}.} The above S-matrix is often found re-expressed using an integral representation (see for instance \cite{KirillovReshetikhin,DoikouNepomechie}):
\begin{eqnarray}
 S= \exp \bigg[ \int_0^\infty \frac{\sinh(v s) \, \sinh\big[s (\Sigma - \frac{1}{2})\big]}{s \, \cosh(\frac{s}{2}) \sinh(\Sigma \, s)} \, ds \, \bigg]. 
\end{eqnarray}

The S-matrix (\ref{sinoGordo}) has also been obtained directly in the spin-chain setting for a spin-zero two-particle state of the spin-$\frac{1}{2}$ XXZ chain in its massless regime \cite{BabujianTsvelik,KirillovReshetikhin,DoikouNepomechie}.   
\medskip

\item {\bf Limit 3. Alternative route to the special functions of the $S$-pairing}

By introducing the variable
\begin{eqnarray}
\delta \equiv \frac{\lambda}{K},
\end{eqnarray}
one can see, for instance, that the Limit 2 above can be equivalently obtained as $\delta \to 0$. Alternatively, \cite{CauxKonnoSorrellWeston} reports a limit where a suitably introduced lattice spacing $\epsilon$ scales to zero\footnote{See also Sect. 6 of \cite{FateevFradkinLukyanovZamolodchikovZamolodchikov}, where, in their conventions, the lattice spacing enters as $J_x \rightarrow \frac{J_x}{\epsilon}$, $J_y \rightarrow \frac{J_y}{\epsilon}$ and $J_z \rightarrow \frac{J_z}{\epsilon}$.} alongside with the parameter $\delta$ such that
$$
\delta \to 0, \qquad \epsilon \to 0, \qquad 2 \, \frac{e^{- \frac{\pi}{\delta}}}{\epsilon} \to M \, \mbox{finite}.
$$ 
The number of spin-chain sites is also taken to be infinite, hence, in this limit, the discrete chain tends to a continuum model, which turns out to be the Sine-Gordon theory \cite{LukyanovTerras}. The parameter $M$ plays the role of the mass entering the particle dispersion relation in the continuum model. The Sine-Gordon spectrum is massive and consists of a soliton, an anti-soliton and a tower of corresponding bound states (the so-called {\it breathers}).

Since the Sine-Gordon limit still involves sending $\delta \to 0$, we expect a similar type of S-matrix as in the case of Limit 2. In fact, the limiting expression in terms of an infinite product of gamma functions famously reproduces (apart from overall factors) the Jost function for the (anti-)soliton and (anti-)soliton scattering in the Sine-Gordon theory \cite{ZamolodchikovZamolodchikov},  or, equivalently, for the excitations of the massive Thirring model \cite{Korepin}, \cite{Coleman}. The S-matrix is given by  
(\ref{sinoGordo}), with the parameter $\Sigma$ now related to the Sine-Gordon coupling constant. 
 
\medskip

The XXZ chain in its disordered regime has been connected to a lattice regularization of Sine-Gordon and Liouville theory in \cite{FaddeevTirkkonen}. A similar relationship between the modular XXZ chain and the lattice Sinh-Gordon theory has been explored in \cite{BytskoTeschner}. 

\end{itemize}

\bigskip

We remark that in \cite{gaddegukovputrovwalls} it was shown that 3d  $\mathcal{N}=2$ solid tori  partition functions satisfy the 
Baxter equation  for the $\mathfrak{sl}(2)$ XXZ spin-chain.
In appendix \ref{fabax}  we offer another derivation of this relation showing that 
the hypergeometric difference equation
satisfied by the 3d holomorphic blocks ${\cal B}^{3d}_\alpha$ can be mapped to the Baxter equation  for the $\mathfrak{sl}(2)$ XXZ spin-chain.

For the sake of completeness, we recall that the very same S-matrix (\ref{sinoGordo}) also features in the scattering of two spin-$\frac{1}{2}$ spin-wave excitations propagating on the {\it antiferromagnetic} XXX spin-chain with arbitrary spin $\Sigma$ representation at each site \cite{Takhtajan,Reshetikhin} (the spin $\Sigma$ entering the formula in a similar way as the coupling constant does in the case of the Sine-Gordon model). Namely, for a spectral parameter $w$ and a singlet-triplet system of excitations, 
\begin{eqnarray}
S \, = \, S_{\frac{1}{2}} \, \cdot \frac{\sinh\big(\frac{\pi}{4\Sigma} (w - i)\big)}{\sinh\big(\frac{\pi}{4\Sigma} (w + i)\big)} \, \, \exp \bigg[- i \int_0^\infty \frac{\sin(w s) \, \sinh\big[s (\Sigma - \frac{1}{2})\big]}{s \, \cosh(\frac{s}{2}) \sinh(\Sigma \, s)} \, ds \, \bigg],
\end{eqnarray}
where $S_{\frac{1}{2}}$ is the S-matrix for spin-$\frac{1}{2}$ particles, related to the central extension of the { $\alg{sl}(2)$} Yangian double \cite{KhoroshkinLebedevPakuliak}:
\begin{eqnarray}
\label{XXX}
S_{\frac{1}{2}} \, = \, \frac{\Gamma(- i \frac{w}{2})\Gamma(\frac{1}{2} + i \frac{w}{2})}{\Gamma(i \frac{w}{2})\Gamma(\frac{1}{2} - i \frac{w}{2})} \, \cdot \, \frac{w - i \, { \mathbb{P}}}{w - i}, \qquad { \mathbb{P}} = \mbox{permutation in $\mathbb{C}^2 \otimes \mathbb{C}^2$}. 
\end{eqnarray}
Let us finally point out that the S-matrix (\ref{XXX}) appears also in connection with the massless limit of the $\alg{su}(2)$ Thirring model \cite{ZamolodchikovZamolodchikovI}. 

\smallskip

\noindent If one proceeds one steps ``downwards", one can take the limit $q \to 1$ of (\ref{XXZ}) and obtain a Jost function written in terms of a ratio of ordinary Gamma functions. This reproduces \cite{FreundZabrodinI} the scattering of two spin-$\frac{1}{2}$ spin-wave ({\it kink}) excitations of the XXX spin-$\frac{1}{2}$ spin-chain \cite{FaddeevTakhtajanI,FaddeevTakhtajan}, {\it i.e.} the triplet part of formula (\ref{XXX})\footnote{ In fact, when the S-matrix (\ref{XXX}) acts on two scattering excitations prepared in a triplet (symmetric) state of $\bf \frac{1}{2} \otimes \frac{1}{2}$, the permutation operator $\mathbb{P}$ produces eigenvalue $1$ and the factor $\frac{w - i \, \mathbb{P}}{w - i}$ equals $1$, leaving the Gamma-function prefactor as a result.}.
This limit corresponds to sending $\gamma \sim \lambda \to 0$ (see appendix \ref{parame}, where $K'_l$ has been sent to a constant in Limit 1),  from which we see that the Hamiltonian reduces to the one of the {\it antiferromagnetic isotropic} spin-chain, for which one can then use the equivalence of spectra $H(\Delta) \leftrightarrow -H(-\Delta)$ \cite{Samaj}.
This in turn produces a similar structure as the mini-superspace Liouville reflection coefficient, which has been obtained in \cite{OlshanetskyRogov} by studying the ordinary Lobachevsky space (see also \cite{GerasimovMarshakovOlshanetskyShatashvili,GerasimovMarshakovMorozovOlshanetskyShatashvili}). 
This is in agreement with the fact that in the $q\to 1$ limit the $id$-CFT reduces to Liouville theory \cite{Nieri:2013yra}.

\smallskip

We remark that the semiclassical  reflection coefficient of the Liouville theory contains { only part of the} gamma functions present in formula (\ref{XXX}).

 This is related to the fact that the reflection coefficient, in the semiclassical limit $b_0 \to 0$, can also be interpreted as the S-matrix for the scattering of a quantum mechanical particle against a static potential\footnote{In some cases it may be necessary to take specific limits on the parameters of the S-matrix.} (mini-superspace approximation { \cite{Zamolodchikov:1995aa,Seiberg}}). In one case the potential is given by the Liouville one, in the other case it is given by the (Calogero-Moser-Sutherland type) potential $\propto \sinh^{-2} x$ \cite{OlshanetskyPerelomov} (see also \cite{Inozemtsev}).

 The other reflection coefficients we have derived in the previous section, {\it i.e.} for the $id$- and $S$-pairing, call for analogous considerations. For the $id$-pairing, producing the Harish-Chandra functions of \cite{OlshanetskyRogov}, and similarly for the $S$-pairing, we again retain a reduced number of the special functions as compared to what characterizes the S-matrices of the integrable (spin-chain) models we discuss in this section. 

\smallskip

Let us finally point out once more that in this whole discussion it is understood that affinization has yet to be performed. It would be very interesting to study what spin-chain picture might arise, if any, when the affinization takes place.

\section*{Acknowledgments}
It is a pleasure to thank Giulio Bonelli, Davide Fioravanti, Domenico Orlando, Vladimir Rubtsov and Maxim Zabzine for discussions{, and Maxim Zabzine and Jian Qiu for sharing a version of their paper in preparation}. 
The work of F.P. is supported by a Marie Curie International Outgoing Fellowship  FP7-PEOPLE-2011-IOF, Project n\textsuperscript{o} 298073 (ERGTB).
A.T. thanks EPSRC for funding under the First Grant project EP/K014412/1 ``Exotic
quantum groups, Lie superalgebras and integrable systems", and the Mathematical Physics group at the University of York for kind hospitality during the preparation of this paper.
The work of F.N.
is partially supported by the EPSRC - EP/K503186/1.

\appendix

\section{Special functions}
In this appendix we describe few of the special functions and identities used in the main text.
\subsection{Bernoulli polynomials}\label{ber}
Throughtout this appendix let us denote by
\be
\vec{\omega}:=(\omega_1,\ldots,\omega_r) \in \mathbb{C}^r
\ee
a vector of $r$ parameters. The multiple Bernoulli polynomials $B_{rr}(z|\vec{\omega})$ up to cubic order are defined by \cite{naru}
\ben
B_{11}(z|\vec{\omega})=&&\frac{z}{\omega_1}-\frac{1}{2}\nn\\
B_{22}(z|\vec{\omega})=&&\frac{z^2}{\omega_1\omega_2}-\frac{\omega_1+\omega_2}{\omega_1\omega_2}z+\frac{\omega_1^2+\omega_2^2+3\omega_1\omega_2}{\omega_1\omega_2}\nn\\
B_{33}(z|\vec{\omega})=&& \frac{z^3}{\omega_1 \omega_2 \omega_3}
- \frac{3\,(\omega_1 + \omega_2 + \omega_3)}
{2\, \omega_1 \omega_2 \omega_3} z^2 
 + \frac{\omega_1^2 + \omega_2^2 + \omega_3^2
+ 3(\omega_1 \omega_2 + \omega_2 \omega_3 + \omega_3 \omega_1)}
{2\, \omega_1 \omega_2 \omega_3} z\nonumber \\ 
&& - \frac{(\omega_1 + \omega_2 + \omega_3)
(\omega_1 \omega_2 + \omega_2 \omega_3 + \omega_3 \omega_1)}
{4\, \omega_1 \omega_2 \omega_3}\,.
\een
If not stated otherwise, we will use the shorthand notation $B_{rr}(z):=B_{rr}(z|\vec{\omega})$.
\subsection{Multiple Gamma and Sine functions}\label{tri}

The Barnes $r$-Gamma function $\Gamma_r(z|\vec{\omega})$ can be defined as the $\zeta$-regularized infinite product \cite{naru}
\be
\Gamma_r(z|\vec{\omega})\sim \prod_{\vec{n}\in \mathbb{Z}^+_0}\frac{1}{\left(z+\vec{\omega}\cdot\vec{n}\right)}.
\ee
When there is no possibility of confusion, we will simply set $\Gamma_r(z):=\Gamma_r(z|\vec{\omega})$.\\

The $r$-Sine function is defined according to \cite{naru}
\be
S_r(z|\vec{\omega})=\frac{\Gamma_r(E_r-z)^{(-1)^r}}{\Gamma_r(z)}
\ee
where we defined $E_r:=\sum_i\omega_i$. We will also denote $S_r(z):=S_r(z|\vec{\omega})$ when there is no confusion. Also, introducing the multiple $q$-shifted factorial
\be
\label{mulq}
\left(z;q_1,\ldots q_r\right):=\prod_{k_1,\ldots, k_r\geq 0}\left(1-zq_1^{k_1}\cdots q_r^{k_r}\right)
\ee
the $r$-sine function has the following product representation ($r\geq 2$) \cite{naru}
\be\label{s3fac}
S_r(z)=e^{(-1)^r\frac{i \pi }{r!}B_{rr}(z)}\prod_{k=1}^r\left(e^{\frac{2\pi i}{\omega_k}z};e^{2\pi i\frac{\omega_1}{\omega_k}},\ldots,e^{2\pi i\frac{\omega_{k-1}}{\omega_k}},e^{2\pi i\frac{\omega_{k+1}}{\omega_k}},\ldots,e^{2\pi i\frac{\omega_{r}}{\omega_k}}\right)
\ee
whenever ${\rm Im}\left( \omega_j/\omega_k\right) \neq 0$ (for $j\neq k$). General useful identities are 
\be
S_r(z)S_r(E_r-z)^{(-1)^r}=1 
\ee
\be
S_r(\lambda z|\lambda\vec{\omega})=S_r(z|\vec{\omega});\quad \lambda \in \mathbb{C}/\{0\}
\ee
\be
\label{prorat}
\frac{S_r(z+\omega_i)}{S_r(z)}=\frac{1}{S_{r-1}(z|\omega_1,\ldots,\omega_{i-1},\omega_{i+1},\ldots,\omega_r)}
\ee
Notice for $r=3$ we can write
\be
S_3(z)=e^{-\frac{i \pi }{3!}B_{33}(z)}\left(e^{\frac{2\pi i}{e_3}z};q,t\right)_1\left(e^{\frac{2\pi i}{e_3}z};q,t\right)_2\left(e^{\frac{2\pi i}{e_3}z};q,t\right)_3
\ee
where $q$, $t$ are expressed via the $e_1$, $e_2$, $e_3$ parameters as described in (\ref{p123}), (\ref{pq}), and it is customary to denote $E=\omega_1+\omega_2+\omega_3$. For $r=2$ it is convenient to introduce the double sine function
\be
s_b(z)=S_2(Q/2-iz|b,b^{-1})\sim \prod_k\frac{n_1\omega_1+n_2\omega_2+Q/2-iz}{n_1\omega_1+n_2\omega_2+Q/2+iz}
\ee
where it is customary to denote $Q=\omega_1+\omega_2$, and it is usually assumed $b=\omega_1=\omega_2^{-1}$.

\subsection{$\Upsilon^\beta$ function}\label{upb}
The $q$-deformed version of the Euler $\Gamma$ function is defined as
\be
\Gamma_q(z):=\frac{(q;q)}{(q^z;q)}(1-q)^{1-z}.
\ee
It has the following classical limit
\be
\Gamma_q(z)\stackrel{q\to 1}{\longrightarrow}\Gamma(z)
\ee
and satisfies the functional relation
\be
\Gamma_q(1+z)=\frac{1-q^z}{1-q}\Gamma_q(z).
\ee\\

A  deformation of the $\Upsilon(z)$ function appearing in Liouville field theory
\be
\Upsilon(z)=\Gamma_2(z|b_0,b_0^{-1})^{-1}\Gamma_2(Q_0-z|b_0,b_0^{-1})^{-1}
\ee
where $Q_0:=b_0+b_0^{-1}$, is the $\Upsilon^\beta(z)$ function defined as the $\zeta$-regularized infinite product
\ben\nn
\Upsilon^\beta(z)
&\sim& \prod_{n_1,n_2\geq0}\sinh\left[\frac{\beta}{2}\left(z+n_1 b_0+n_2 b_0^{-1}\right)\right]\sinh\left[\frac{\beta}{2}\left(Q_0-z+n_1b_0+n_2b_0^{-1}\right)\right]\,\nn\\
&&\nn\\
\label{sym}&&\sim\left(e^{\beta z};e^{\beta /b_0},e^{\beta b_0}\right)\left(e^{-\beta z};e^{-\beta/b_0},e^{-\beta b_0}\right)
\een
By a suitable regularization, important defining properties are 
\be
\Upsilon^\beta(z)=\Upsilon^\beta(Q_0-z)
\ee
\be
\frac{\Upsilon^\beta(z+b_0^{\pm 1})}{\Upsilon^\beta(z)}\sim \frac{\left(e^{\beta(b_0^{\mp 1}- z)};e^{\beta b_0^{\mp 1}}\right)}{\left(e^{\beta z};e^{\beta b_0^{\mp 1}}\right)}=\frac{1}{\left(e^{\beta z};e^{\beta b_0^{\mp 1}}\right)\left(e^{-\beta z};e^{-\beta b_0^{\mp 1}}\right)}.
\ee
Using the expressions (\ref{pq}) and (\ref{idqt}), we can finally write 
\ben\label{idfac}
\Upsilon^\beta(z)&&\sim \left(e^{\frac{2\pi i}{e_3} z};q,t\right)_1\left(e^{\frac{2\pi i}{e_3} z};q,t\right)_2\,.
\een

\subsection{Jacobi Theta and elliptic Gamma functions}\label{tega}
The Jacobi $\Theta$ function is defined by \cite{grahman}
\be
\Theta(z;\tau)=\left(e^{2\pi i z};e^{2\pi i\tau}\right)\left(e^{2\pi i\tau}e^{-2\pi i z};e^{2\pi i\tau}\right)
\ee
and satisfies the functional relation
\be
\frac{\Theta(\tau+z;\tau)}{\Theta(z;\tau)}=-e^{-2\pi i z},
\ee
or more generally
\be
\frac{\Theta\left(n\tau+z;\tau\right)}{\Theta\left(z;\tau\right)}=\left[-e^{2\pi i z}\left(e^{2\pi i\tau}\right)^{\frac{n-1}{2}}\right]^{-n}
\ee
for $n\in\mathbb{Z}^+_0$. Another relevant property is \cite{naru}
\be
\Theta\left(\frac{z}{\omega_1};\frac{\omega_2}{\omega_1}\right)\Theta\left(\frac{z}{\omega_2};\frac{\omega_1}{\omega_2}\right)=e^{-i\pi B_{22}(z)}.
\ee
\\

The elliptic Gamma function $\Gamma_{q,t}$ is defined by \cite{grahman}
\be
\Gamma_{q,t}(z)=\frac{(qt\;e^{-2\pi i z};q,t)}{(e^{2\pi i z};q,t)};\quad q=e^{2\pi i\tau};\quad t=e^{2\pi i\sigma}
\ee
and satisfies the functional relations
\be
\frac{\Gamma_{q,t}(\tau+z)}{\Gamma_{q,t}(z)}=\Theta(z;\sigma);\quad \frac{\Gamma_{q,t}(\sigma+z)}{\Gamma_{q,t}(z)}=\Theta(z;\tau),
\ee
or more generally
\ben
\frac{\Gamma_{q,t}\left(n\tau+z\right)}{\Gamma_{q,t}\left(z\right)}&=&\prod_{k=1}^{n}\Theta\left((k-1)\tau+z;\sigma\right);\quad \frac{\Gamma_{q,t}\left(n\sigma+z\right)}{\Gamma_{q,t}\left(z\right)}=\prod_{k=1}^{n}\Theta\left((k-1)\sigma+z;\tau\right)\nn\\
&&\nn\\
\frac{\Gamma_{q,t}\left(n_1\tau+n_2\sigma+z\right)}{\Gamma_{q,t}\left(z\right)}
&=&\left[-e^{2\pi i z}\left(e^{2\pi i \tau}\right)^{\frac{n_1-1}{2}}\left(e^{2\pi i \sigma}\right)^{\frac{n_2-1}{2}}\right]^{-n_1n_2}\times\nn\\
&&\times\prod_{k=1}^{n_1}\Theta\left((k-1)\tau+z;\sigma\right)\prod_{j=1}^{n_2}\Theta\left((j-1)\sigma+z;\tau\right)
\een
for $n_1$, $n_2\in\mathbb{Z}^+_0$. Other relevant properties are \cite{fv}
\be
\Gamma_{q,t}\left(\frac{z}{e_3}\right)_1\Gamma_{q,t}\left(\frac{z}{e_3}\right)_2\Gamma_{q,t}\left(\frac{z}{e_3}\right)_3=e^{-\frac{i\pi}{3}B_{33}(z)}
\ee
where $q$, $t$ are expressed via the $e_1$, $e_2$, $e_3$ parameters as described in (\ref{p123}), (\ref{pq})
and 
\be
\label{gamma unit}
\Gamma_{q,t}\left(\frac{z}{e_3}\right)\Gamma_{q,t}\left(\frac{e_1+e_2-z}{e_3}\right)=1\,.
\ee
\section{Instanton partition function degeneration}\label{instdeg}

The $\mathbb{R}^4\times S^1$ instanton partition function (with rescaled parameters and equivariant parameters $\epsilon_{1,2}=\frac{e_{1,2}}{e_3}$) for the  $SU(2)$ SCQCD  is given by  \cite{Nekrasov:2002qd, Nekrasov:2003rj}

\ben\label{sin}
{\cal Z}^{\mathbb{R}^4\times S^1}_{\rm inst}\left(\frac{\vec{a}}{e_3},\frac{\vec{m}}{e_3}; 
 \frac{e_1}{e_3},\frac{e_2}{e_3}\right)&=&\sum_{\vec{Y}} z^{|\vec{Y}|}
\frac{F_{\vec{Y}}(\vec{a},\vec{m})}{V_{\vec{Y}}(\vec{a})}\,,  \qquad {\rm with } \qquad  z=e^{\frac{2\pi i}{\tilde g^2 e_3}}\,,
\een
where $\vec{Y}=(Y^1,Y^2)$ is a vector of Young diagrams, $\vec{a}=(a_1,a_2)=(a,-a)$  parametrizes the Coulomb branch and $\vec{m}=(m_1,\ldots, m_4)$ are the masses of the four fundamental hypermultiplets. $F_{\vec{Y}}(\vec{a}, \vec{m})$ and $V_{\vec{Y}}(\vec{a})$, the contribution of the fundamental hypermultiplets and of the vector multiplet, are given by:
\ben
F_{\vec{Y}}(\vec{a},\vec{m})&=&\prod_{m=1}^2\prod_{(i,j)\in Y^m}\prod_{f=1}^4\sinh \frac{i\pi}{e_3}[a_m+m_f+(j-1)e_1+(i-1)e_2]\,,
\label{dt}
\een
\ben
\label{vt}
V_{\vec{Y}}(\vec{a})&=&\prod_{m,n=1}^2\prod_{(i,j)\in Y^m}\sinh \frac{i\pi}{e_3}[a_m-a_n-e_1(Y^n_i-j)+e_2(Y^{m T}_j-i+1)]\times\nn\\
&&\phantom{\prod_{m,n=1}^2\prod_{(i,j)\in Y^m}}\times \sinh \frac{i\pi}{e_3}[a_m-a_n-e_1(Y^n_i-j+1)+e_2(Y^{m T}_j-i)]\,,
\een
where   $Y^m_i$ is the length of the $i$-th row of $Y^m$.   

\subsubsection*{Trivial degeneration} 
Now suppose that $m_1+m_2=-e_3$ and let us evaluate the partition function (\ref{sin}) for:
\be
a_1=m_{1}\,,\quad a_2=m_{2}+e_3\,.
\ee 
We first  notice that a shift by $e_3$
in (\ref{dt}), (\ref{vt}) has a trivial  effect, since everything is rescaled by $e_3$.
It is then easy to see that the only non-zero contribution to $F_{\vec{Y}}$ comes from empty Young tableaux $Y^1=Y^2=\emptyset$. 
This gives the trivial degeneration 
\ben
{\cal Z}^{\mathbb{R}^4\times S^1}_{\text{inst}}\quad\xrightarrow[(a_1,a_2)\rightarrow(m_{1},m_2+e_3)]{}\quad 1\,.
\een

\subsubsection*{Hypergeometric degeneration} 
Now suppose that $m_1+m_2=-e_1$ and let us evaluate the partition function (\ref{sin}) for:
\be
a_1=m_{1}\,,\quad a_2=m_{2}+e_1\,.
\ee 
In this case a shift by $e_1$ has a non-trivial effect, and inspecting   $F_{\vec{Y}}$, we discover we can fill in a column in $Y^1$. So, besides $(Y^1,Y^2)=(\emptyset,\emptyset)$, we get non-vanishing contributions from  $(Y^1,Y^2)=(1^n,\emptyset)$:
\ben
&&\!\!\!\!\!\!\!\!\!\!\!\!\!\!\!\!\!\!\!\!
F_{\vec{Y}}(\vec{a},\vec{m})~ \xrightarrow[(a_1,a_2)\rightarrow(m_{1},m_2+e_1)]{}
F_{1^n,\emptyset}= \prod_{k=1}^n\prod_{f=1}^4\sinh\frac{i\pi}{e_3}[m_{1}+m_f+(k-1)e_2]\,,\\
&&\!\!\!\!\!\!\!\!\!\!\!\!\!\!\!\!\!\!\!\!
V_{\vec{Y}}(\vec{a},\vec{m})~\xrightarrow[(a_1,a_2)\rightarrow(m_{1},m_2+e_1)]{}V_{1^n,\emptyset}=
 \prod_{k=1}^n\sinh\frac{i\pi}{e_3}[2m_{1}+e_1+(n-k+1)e_2]\times\nn\\
&&\times\sinh\frac{i\pi}{e_3}[2m_{1}+(n-k)e_2]\sinh\frac{i\pi}{e_3}[(n-k+1)e_2]\sinh\frac{i\pi}{e_3}[-e_1+(n-k)e_2]\,.
\een
We then simplify the ratio:\footnote{We use $m_1+m_2=-e_1$, $\prod_{k=1}^n f(k)=\prod_{k=1}^n f(n-k+1)$.}

\ben
\frac{F_{1^n,\emptyset}}{V_{1^n,\emptyset}}&=&\prod_{k=1}^n\frac{\sinh\frac{i\pi}{e_3}[m_{1}+m_{3}+(k-1)e_2]\sinh\frac{i\pi}{e_3}[m_{1}+m_{4}+(k-1)e_2]}
{\sinh\frac{i\pi}{e_3}[m_{1}-m_{2}+ke_2]\sinh\frac{i\pi}{e_3}[ke_2]}\nn\\
&=&e^{n\frac{i\pi}{e_3}[\sum_f2e_2-m_{f}]}\times
\prod_{k=0}^{n-1}\frac{(1-e^{\frac{2i\pi}{e_3}[m_{1}+m_{3}+ke_2]})(1-e^{\frac{2i\pi}{e_3}[m_{1}+m_{4}+ke_2]})}
{(1-e^{\frac{2i\pi}{e_3}[m_{1}-m_{2}+(k+1)e_2]})(1-e^{\frac{2i\pi}{e_3}(k+1)e_2})}\nn\\
&=&e^{n\frac{i\pi}{e_3}[\sum_f2e_2-m_{f}]}\times 
\frac{(e^{\frac{2i\pi}{e_3}[m_{1}+m_{3}]};e^{2\pi i\frac{e_2}{e_3}})_n(e^{\frac{2i\pi}{e_3}[m_{1}+m_{4}]};e^{2\pi i\frac{e_2}{e_3}})_n}
{(e^{\frac{2i\pi}{e_3}[m_{1}-m_{2}+e_2]};e^{2\pi i\frac{e_2}{e_3}})_n(e^{2\pi i\frac{e_2}{e_3}};e^{2\pi i\frac{e_2}{e_3}})_n}\,,\qquad\qquad
\een
and finally obtain:
\ben
\label{hyperdeg}
{\cal Z}^{\mathbb{R}^4\times S^1}_{\text{inst}}\left(\frac{\vec{a}}{e_3};\frac{\vec {m}}{e_3}; 
 \frac{e_1}{e_3},\frac{e_2}{e_3}\right)\quad \xrightarrow[(a_1,a_2)\rightarrow(m_{1},m_2+e_1)]{}
~ \sum_{n\geq 0}~\frac{F_{1^n,\emptyset}}{V_{1^n,\emptyset}} z^n= \phantom{|}_2\Phi_1(A,B;C,q;u)\,,
\een
where
\ben\nn
A&=&e^{\frac{2i\pi}{e_3}[m_{1}+m_{3}]}\,,\quad B=e^{\frac{2i\pi}{e_3}[m_{1}+m_{4}]}\,,
\quad C=e^{\frac{2i\pi}{e_3}[m_{1}-m_{2}+e_2]}\,, \quad q= e^{2\pi i\frac{e_2 }{e_3}}\,,\\ \label{par}
r&=& e^{-\frac{2\pi i}{e_3}\sum_fm_{f}}=e^{-2\pi i\frac{e_2}{e_3}}CB^{-1}A^{-1}\,,\quad u=e^{2\pi i\frac{e_2}{e_3}}r^{1/2}z\,,
\een
and we have introduced the  $q$-hypergeometric series $\phantom{|}_2\Phi_1(A,B;C,q;u)$ defined by
\be\label{qhyper21}
\phantom{|}_2\Phi_1(A,B;C,q;u)=\sum_{k\geq 0}\frac{ (A;q)_k (B;q)_k }{(q;q)_k (C;q)_k}u^k\,.
\ee\\

It is also easy to see the condition $m_1+m_2=-e_2$ yields the same result with $e_1\leftrightarrow e_2$,
in this case the non-empty tableaux will be $Y^2$ where we can fill a row $Y^2=n=(1^n)^T$:
\ben
\label{hyperdeg2}
{\cal Z}^{\mathbb{R}^4\times S^1}_{\text{inst}}\left(\frac{\vec{a}}{e_3};\frac{\vec {m}}{e_3}; 
 \frac{e_1}{e_3},\frac{e_2}{e_3}\right)\quad \xrightarrow[(a_1,a_2)\rightarrow(m_{1},m_2+e_2)]{}
~ \sum_{n\geq 0}~\frac{F_{n,\emptyset}}{V_{n,\emptyset}} z^n= \phantom{|}_2\Phi_1(A,B;C,\tilde q;\tilde u)\,.
\een
where the tilde symbol means $e_1\leftrightarrow e_2$.

\subsubsection*{Hook degeneration  }

Now suppose that $m_1+m_2=-\vec n\cdot \vec e$ and let us evaluate the partition function (\ref{sin}) for:
\be
a_1=m_1+(\vec n-\vec p)\cdot \vec e \,, \quad a_2=m_2+\vec p \cdot \vec e \,,
\ee 
with 
\be
p_k\in\{0,1,\ldots,n_k\}\,.
\ee

Inspecting  $F_{\vec{Y}}$,  we observe that 
for fixed $p_1$, $p_2$, $p_3$, we get a zero from the box $(i,j)=(p_2+1,p_1+1)$ in $Y^1$, and the box $(i,j)=(n_2-p_2+1,n_1-p_1+1)$ in $Y^2$. Therefore, non--vanishing contributions are from $(Y^1,Y^2)=(\emptyset,\emptyset)$ and hook shaped tableaux $(Y^1,Y^2)=((p_2,p_1),(n_2-p_2,n_1-p_1))$.
The residue at this point is given by:
\ben
&&F_{\vec{Y}}(\vec{a},\vec{m})
~\xrightarrow[(a_1,a_2)\rightarrow(m_{1}+(\vec n-\vec p)\cdot \vec e,m_2+\vec p\cdot \vec e)]{}
F_{p_2,p_1}=
\nn\\
&&
\prod_{f=3,4}\prod_{(i,j)\in Y^1}\sinh \frac{i\pi}{e_3}[m_1+(\vec n-\vec p)\cdot \vec e+m_f+(j-1)e_1+(i-1)e_2]\times\nn\\
&&\times \prod_{f=3,4}\prod_{(i,j)\in Y^2}\sinh \frac{i\pi}{e_3}[m_2+\vec p\cdot \vec e+m_f+(j-1)e_1+(i-1)e_2]\times\nn\\
&&\times \prod_{(i,j)\in Y^1}\sinh \frac{i\pi}{e_3}[-\vec p\cdot \vec e+(j-1)e_1+(i-1)e_2]
\times\nn\\
&&\times \prod_{(i,j)\in Y^2}\sinh \frac{i\pi}{e_3}[-(\vec n-\vec p)\cdot \vec e+(j-1)e_1+(i-1)e_2]\times\nn\\
&&\times\prod_{(i,j)\in Y^1}\sinh \frac{i\pi}{e_3}[2m_1+(\vec n-\vec p)\cdot \vec e+(j-1)e_1+(i-1)e_2]\times\nn\\
\nn\\
&&\times\prod_{(i,j)\in Y^2}
\sinh \frac{i\pi}{e_3}[2m_2+\vec p\cdot \vec e+(j-1)e_1+(i-1)e_2]\,,\nn\\
&&
\label{morte1}
\een
\ben
&&V_{\vec{Y}}(\vec{a},\vec{m})
~\xrightarrow[(a_1,a_2)\rightarrow(m_{1}+(\vec n-\vec p)\cdot \vec e,m_2+\vec p\cdot \vec e)]{}
V_{p_2,p_1}=
\nn\\
&&\times 
\prod_{(i,j)\in Y^1}\sinh \frac{i\pi}{e_3}[-e_1(Y^1_i-j)+e_2(Y^{1 T}_j-i+1)]
\sinh \frac{i\pi}{e_3}[-e_1(Y^1_i-j+1)+e_2(Y^{1 T}_j-i)]
\times\nn\\
&&\times \prod_{(i,j)\in Y^2}\sinh \frac{i\pi}{e_3}[-e_1(Y^2_i-j)+e_2(Y^{2 T}_j-i+1)]\sinh \frac{i\pi}{e_3}[-e_1(Y^2_i-j+1)+e_2(Y^{2 T}_j-i)]\times\nn\\
&&\times\prod_{(i,j)\in Y^1}\sinh \frac{i\pi}{e_3}[2m_1+2(\vec n-\vec p)\cdot \vec e-e_1(Y^2_i-j)+e_2(Y^{1 T}_j-i+1)]\times\nn\\
&&\phantom{\prod_{(i,j)\in Y^1}}\times\sinh \frac{i\pi}{e_3}[2m_1+2(\vec n-\vec p)\cdot \vec e-e_1(Y^2_i-j+1)+e_2(Y^{1 T}_j-i)]\times\nn\\
&&\times\prod_{(i,j)\in Y^2}\sinh \frac{i\pi}{e_3}[2m_2+2\vec p\cdot \vec e-e_1(Y^1_i-j)+e_2(Y^{2 T}_j-i+1)]\times\nn\\
&&\phantom{\prod_{(i,j)\in Y^2}}\times\sinh \frac{i\pi}{e_3}[2m_2+2\vec p\cdot \vec e-e_1(Y^1_i-j+1)+e_2(Y^{2 T}_j-i)]\,.
\label{morte2}
\een

\subsection{Classical term}
The classical term, up to factors independent of $a$ is,
\be\mathcal{Z}^{\rm cl}=\prod_{i=1,2}
\frac{\Gamma_{q,t}\left(\frac{a_i+1/g^2-\sum_fm_f/2+\kappa}{e_3}\right)}{\Gamma_{q,t}\left(\frac{a_i+\kappa}{e_3}\right)}
\ee
when evaluated at $a_1=m_1+(n-p)\cdot e$, $a_2=m_2+\vec p\cdot \vec e$ yields
\be
\label{degcl1}
\mathcal{Z}^{\rm cl}_{p_2,p_1}=
\frac{\Gamma_{q,t}\left(\frac{m_1+1/g^2-\sum_fm_f/2+\kappa+(\vec n-\vec p)\cdot \vec e}{e_3}\right)}{\Gamma_{q,t}\left(\frac{m_1+\kappa+(\vec n-\vec p)\cdot \vec e}{e_3}\right)}
\frac{\Gamma_{q,t}\left(\frac{m_2+1/g^2-\sum_fm_f/2+\kappa+\vec p\cdot \vec e}{e_3}\right)}{\Gamma_{q,t}\left(\frac{m_2+\kappa+\vec p\cdot \vec e}{e_3}\right)}\,.
\ee
Multiplying (\ref{degcl1}) by
\be
\frac{1}{\mathcal{Z}^{\rm cl}_{0,0}|_{(n_2,n_1)=(0,0)}}=\frac{\Gamma_{q,t}\left(\frac{m_1+\kappa}{e_3}\right)\Gamma_{q,t}\left(\frac{m_2+\kappa}{e_3}\right)}{\Gamma_{q,t}\left(\frac{m_1+1/g^2-\sum_fm_f/2+\kappa}{e_3}\right)\Gamma_{q,t}\left(\frac{m_2+1/g^2-\sum_fm_f/2+\kappa}{e_3}\right)}
\ee
we may rewrite all in terms of $\Theta$'s 
\ben
\label{degcl2}
\mathcal{Z}^{\rm cl}_{p_2,p_1}&\propto&
\left(-e^{\frac{2\pi i}{e_3}(1/g^2-\sum_fm_f/2)}\right)^{-(n_1-p_1)(n_2-p_2)-p_1p_2}\times\nn\\
&&\times\prod_{k=1}^{n_1-p_1}\frac{\Theta\left(\frac{(k-1)e_1+m_1+1/g^2-\sum_fm_f/2+\kappa}{e_3};\frac{e_2}{e_3}\right)}{\Theta\left(\frac{(k-1)e_1+m_1+\kappa}{e_3};\frac{e_2}{e_3}\right)}
\prod_{j=1}^{n_2-p_2}\frac{\Theta\left(\frac{(j-1)e_2+m_1+1/g^2-\sum_fm_f/2+\kappa}{e_3};\frac{e_1}{e_3}\right)}{\Theta\left(\frac{(j-1)e_2+m_1+\kappa}{e_3};\frac{e_1}{e_2}\right)}\nn\\
&&\times \prod_{k=1}^{p_1}\frac{\Theta\left(\frac{(k-1)e_1+m_2+1/g^2-\sum_fm_f/2+\kappa}{e_3};\frac{e_2}{e_3}\right)}{\Theta\left(\frac{(k-1)e_1+m_2+\kappa}{e_3};\frac{e_2}{e_3}\right)}
\prod_{j=1}^{p_2}\frac{\Theta\left(\frac{(j-1)e_2+m_2+1/g^2-\sum_fm_f/2+\kappa}{e_3};\frac{e_1}{e_3}\right)}{\Theta\left(\frac{(j-1)e_2+m_2+\kappa}{e_3};\frac{e_1}{e_3}\right)}\,.\nn\\
&&
\een

\section{Transfer matrices and Baxter operators}\label{BaxtOp}
We will briefly recall here the notion of Baxter operator for the simplest case of the homogenoeous Heisenberg (XXX) spin-chain. The main references we follow for this are \cite{BazhanovLukowskiMeneghelliStaudacher} and \cite{FaddeevH}.

As anticipated in the discussion following (\ref{XXX}), the isotropic Heisenberg chain (which we now take to be ferromagnetic for simplicity, this not affecting the main conclusions of this appendix) reads
\begin{eqnarray}
\label{HamXXX}
H = \sum_{n=1}^N \, \big(1 \, - \, \sigma_n^x \, \sigma_{n+1}^x - \, \sigma_n^y \, \sigma_{n+1}^y - \, \sigma_n^z \, \sigma_{n+1}^z\big),
\end{eqnarray}
where we have set $J_x = J_y = J_z = 4$ and added a constant w.r.t. (\ref{HXYZ}), to normalize to zero the energy of the ferromagnetic ground state. All sites carry a fundamental representation of $\alg{su}(2)$.

In the framework of the so-called {\it algebraic Bethe ansatz} (ABA), one constructs a Lax matrix acting as a two-by-two matrix on an auxiliary space also carrying the fundamental representation of $\alg{su}(2)$, with matrix-entries acting on the $n$-th site of the chain:
\begin{eqnarray}
L_{a,n} (\rho) = \begin{pmatrix}\rho + \frac{i}{2} \sigma_n^z& \frac{i}{2} \sigma_n^- \\ \frac{i}{2} \sigma_n^+ & \rho - \frac{i}{2} \sigma_n^z
\end{pmatrix},
\end{eqnarray}
where $\sigma_n^\pm = \sigma_n^x \pm i \sigma_n^y$, and $\rho$ is an auxiliary variable.

One then construct the so-called {\it monodromy matrix} as
\begin{eqnarray}
\label{ABCD}
T_a (\rho) = L_{a,1} \cdot \, ... \, \cdot L_{a,N} \, = \, \begin{pmatrix}A(\rho)&B(\rho)\\C(\rho)&D(\rho)
\end{pmatrix}
\end{eqnarray}
($\cdot$ denoting matrix multiplication in the auxiliary space),
acting on the auxiliary space as a two-by-two matrix, with entries $A(\rho)$, $B(\rho)$, $C(\rho)$ and $D(\rho)$ acting on the whole spin-chain. Because of the relation
\begin{eqnarray}
R_{a_1,a_2} (\rho_1 - \rho_2) \, L_{a_1,n_1} (\rho_1) \, L_{a_2,n_2} (\rho_2)\, = \, L_{a_2,n_2} (\rho_2)\, L_{a_1,n_1} (\rho_1)\, R_{a_1,a_2} (\rho_1 - \rho_2),
\end{eqnarray}
with the $\alg{su}(2)$ Yangian R-matrix given by
$$
R_{a_1,a_2} (\rho) = \rho \, I_{a_1,a_2} + i \, { \mathbb{P}_{a_1,a_2}}
$$
($I_{a_1,a_2}$ being the identity operator, { $\mathbb{P}_{a_1,a_2}$} the permutation operator on the two isomorphic auxiliary spaces $a_1$ and $a_2$), one deduces
\begin{eqnarray}
\label{RTT}
R_{a_1,a_2} (\rho_1 - \rho_2) \, T_{a_1} (\rho_1)\, T_{a_2} (\rho_1)\, = \, T_{a_2} (\rho_1)\, T_{a_1} (\rho_1)\, R_{a_1,a_2} (\rho_1 - \rho_2).
\end{eqnarray}
In turn, by taking the trace $tr_{a1} \otimes tr_{a_2}$ on both sides of (\ref{RTT}), one obtains that the {\it transfer matrix} $T(\rho) \equiv tr \, T_a (\rho) = A(\rho) + D(\rho)$ commute for two arbitrary values of the auxiliary variable:
\begin{eqnarray}
\label{being}
[T(\rho), T(\rho')]=0.
\end{eqnarray}
Being $T(\rho)$, by inspection, an order $N$ polynomial in $\rho$ (with the highest-power coefficient equal to 1), (\ref{being}) implies that $T(\rho)$ generates $N$ non-trivial independent commuting operators, given by the corresponding polynomial coefficients. The original Hamiltonian (\ref{HamXXX}) is obtained as
$$
H = - 2 i \, \bigg[ \frac{d}{d \rho} \ln T(\rho)\bigg]_{\, \rho = \frac{i}{2}},
$$
hence all the $N$ commuting operators generated by $T(\rho)$ commute with $H$, and are therefore integrals of motion. This determines the complete integrability of the problem.

The relations (\ref{RTT}), sometimes called {\it RTT relations}, allows to find the simultaneous eigenvectors of the Hamiltonian and of all the commuting charges, by utilising $B (\rho)$ as a creation operator. One first constructs the family of vectors
\begin{eqnarray}
\label{BBBB}
|\Psi (\rho_1,...,\rho_M) \rangle = B (\rho_1) ... B (\rho_M) \, |vac\rangle,
\end{eqnarray}
with the {\it pseudo-vacuum} $|vac\rangle$ being any highest-weight $T(\rho)$-eigenstate in the representation { ${\bf \frac{1}{2}}^{\otimes N}$} in which the spin-chain transforms\footnote{The pseudo-vacuum is typically chosen as the ``all-spin-up" ferromagnetic vacuum, whether - as in the case of this appendix - or not that is the true ground state of $H$.}. The vectors (\ref{BBBB}) are not automatically eigenstates of $T(\rho)$ because of some unwanted terms one obtains when acting with $T(\rho)$ on such vectors, which are not proportional to the vector themselves. These unwanted terms can be cancelled by imposing the following system of $M$ {\it Bethe equations}:
\begin{eqnarray}
\label{BAE}
\Bigg(\frac{\rho_j + \frac{i}{2}}{\rho_j - \frac{i}{2}}\Bigg)^N \, = \, \prod_{m\neq j}^M \frac{\rho_j - \rho_m + i}{\rho_j - \rho_m - i}.
\end{eqnarray}
These equations coincide with the quantization condition for the momenta $p_m$, parameterised according to $e^{i p_m} = \frac{\rho_m + \frac{i}{2}}{\rho_m - \frac{i}{2}}, m=1,...,M$, of $M$ excitations propagating along the chain and collectively described by a scattering-type wave-function ({\it coordinate Bethe ansatz}), as originally found by Bethe \cite{Bethe}. Upon imposing (\ref{BAE}), the vectors (\ref{BBBB}) become eigenstates of $T(\rho)$ and, in particular, they have an energy eigenvalue equal to the sum of the single-particle dispersion relations $\sum_{m=1}^M E(p_m)$. In turn, $E(p)$ is given by the $H$-eigenvalue for $M=1$.

\smallskip

In order to introduce the Baxter operator, one can re-write the Bethe equations (\ref{BAE}) as
\begin{eqnarray}
\label{b9}
\alpha^N (\rho_j) \, {q_M}(\rho_j - i) \, = \, \delta^N (\rho_j) \, { q_M}(\rho_j + i), \qquad j=1,...,M,
\end{eqnarray}
where
\begin{eqnarray}
\alpha(\rho) = \rho + \frac{i}{2}, \qquad \delta(\rho) = \rho - \frac{i}{2}, \qquad
{ q_M}(\rho) = \prod_{m=1}^M (\rho - \rho_m).
\end{eqnarray}
From this it is clear, by direct subsititution and by using the Bethe equations in the form (\ref{b9}), that the quantity ${ U_M}(\rho) := \alpha^N (\rho) \, { q_M}(\rho - i) \, + \, \delta^N (\rho) \, { q_M}(\rho + i)$ vanishes for $\rho=\rho_j, \forall j=1,...,M$, and it is therefore proportional to ${ q_M}(\rho)$. Since it is a polynomial of order $N+M$ in $\rho$, ${ U_M}(\rho)$ must also be proportional to a polynomial with $N$ roots, which turns out to be the eigenvalue ${t_M}(u)$ of $T(u)$ on the eigenstate (\ref{BBBB}). One can then eventually write
{\begin{eqnarray}
\label{Baxtere}
t_M(\rho) \, q_M(\rho) \, = \, \alpha^N (\rho) \, q_M(\rho - i) \, + \, \delta^N (\rho) \, q_M(\rho + i).
\end{eqnarray}}
Starting from this equation, Baxter \cite{BaxterP} postulated the existence of an operator $Q(u)$, diagonal in the same basis (\ref{BBBB}) as (hence commuting with) $T(u)$, this time with eigenvalue ${ q_M}(u)$. The equation (\ref{Baxtere}) can then be interpreted as a relation between eigenvalues, projection on eigenstates of in fact a deeper operatorial relation connecting the operators themselves, called the {\it Baxter equation}:
\begin{eqnarray}
\label{Baxtereq}
T(\rho) \, Q(\rho) \, = \, \alpha^N (\rho) \, Q(\rho - i) \, + \, \delta^N (\rho) \, Q(\rho + i).
\end{eqnarray}
Notice that, due to the highest-weigth property of $|vac\rangle$, the coefficient functions $\alpha^N (\rho)$ and $\delta^N (\rho)$ equal the pseudo-vacuum eigenvalues of the operators $A(\rho)$ and $D(\rho)$ in (\ref{ABCD}), respectively.

The idea of Baxter is that the problem of diagonalising the spin-chain Hamiltonian can be equivalently reformulated in the one of constructing an operator $Q(u)$ satisfying (\ref{Baxtereq}) with certain analiticity requirements. This turns out to provide a more efficient method than the algebraic Bethe ansatz itself, as it also works in more complicated cases where the ABA does not apply.

\section{Relationships amongst the spin-chain parameters}{\label{parame} 
In order to make contact with the spin-chain parameterization, one might find the following dictionary useful. If we call $q_B$ the parameter called $q$ in \cite{BaxterBook} (cf. formulae (8.13.52), (15.5.9) and (15.5.12), for instance), and ($q_{RT}$, $w_{RT}$) the parameters called ($q$, $w$) respectively, in \cite{RicheyTracy}, we conclude  
$$
q_B = q_{RT}^2 = e^{2 i \pi \tau}, \qquad \gamma = i \pi \, w_{RT}.
$$
Let us now call $q_{FZ}$ what is called $q$ in \cite{FreundZabrodin}, and which coincides with $q_B$. The variables ($z$, $q_{RT}$) in \cite{RicheyTracy} are the same as ($z$, $q_{FZ}^{\frac{1}{2}}$) of \cite{FreundZabrodin}, where an equivalent expression to (\ref{XYZ}) is obtained by analysing the formula for the partition function in \cite{RicheyTracy}, see also \cite{FreundZabrodinExc} and \cite{BaxterBook}. Notice that $q_B$ should not be confused with the $q$ appearing in (\ref{q}), where we use the terminology of \cite{FreundZabrodinIV}. The parameter $\tau$ is the same in all three references \cite{FreundZabrodinIV}, \cite{RicheyTracy} and \cite{FreundZabrodin}. Finally, the integer $n$ in \cite{RicheyTracy,FreundZabrodin} should be set equal to $2$ to recover the XYZ chain from Baxter's $\mathbb{Z}_n$ models.

From the spin-chain perspective, the fact that $r$ should be kept fixed in Limit 2 (\ref{gammante}) can be motivated as follows. Formula (7.8) in \cite{BaxterP} sets $\lambda_B = \frac{\pi \, \zeta}{K'_{l}}$, where the subscript $B$ in $\lambda_B$ indicates that this is the $\lambda$ variable used in that paper, and $K'_{l}$ is the complete elliptic integral of the first kind with modulus $l'$. By comparing the partition functions presented in \cite{RicheyTracy} and in \cite{BaxterP}, one deduces that it must be $\lambda_{RT} = \frac{\lambda_{B}}{2 \pi}$, which is consistent with $w_{RT} = - i \, \lambda_{RT}$. The modulus $k$ (which we find to be the same in \cite{Baxter} as in \cite{BaxterBook}, hence the same as we are using here) is related to $l'$ as in formula (5.5) of \cite{BaxterP}, namely $k = \frac{1-l}{1+l}$, $l' = \sqrt{1 - l^2}$ being defined below formula (7.8) in \cite{BaxterP}. This means that, when $k \to 1$, the integral $K'_{l}$ diverges as $\log k'$, sending $\gamma = i \pi \, w_{RT} = \pi \, \lambda_{RT} = \frac{\lambda_B}{2}$ to zero as $\frac{\pi}{2}\frac{\zeta}{K'_{l}} \sim \frac{\mbox{const}}{\log k'}$, for ``$\mbox{const}$" a constant if $\zeta$ is kept fixed. At the same time, $\tau$ goes to zero with the same speed as $\frac{i}{2} \frac{K'}{K} \sim \frac{\mbox{const'}}{\log k'}$, (with ``$\mbox{const}' \,$" another constant), which confirms that $r$ will tend to a constant in the limit. On the other hand, by comparing \cite{BaxterP} and \cite{BaxterBook}, we conclude that the $\lambda$ used in formula (10.4.21) of \cite{BaxterBook}, coincident with the $\lambda$ we use here, should be proportional to $\zeta$ by some proportionality factor which stays finite in the limit, hence our $\lambda$ can be kept finite in the expression for the limiting Hamiltonian.

\section{XXZ Baxter equation and 3d blocks}\label{fabax}

The study of reflection coefficients indicates the partition functions/$q$-CFT correlators we have been focusing on are related to a class of integrable systems via an underlying infinite dimensional $\mathfrak{sl}(2)$ symmetry algebra ($q$-deformed and/or affine), XXZ spin-chains being particular representatives. The aim of this section is to explore further the connection between gauge/$q$-Virasoro theories and integrable systems.
We will provide an alternative derivation of the result obtained in   \cite{gaddegukovputrovwalls} that  the  $q$-difference equation 
satisfied by the 3d blocks can be mapped to the Baxter equation of the XXZ 
spin-chain.\\

To begin with, let us remind that the two 3d holomorphic blocks $\mathcal{B}^{\rm 3d}_{1,2}$ for the $U(1)$ theory with
two chirals of charge plus and two chirals of charge minus (see for example section 2.2 in 
\cite{Nieri:2013yra}) are proportional to the two solutions of the $q$-hypergeometric equation 
\be
\mathcal{B}^{\rm 3d}_i(u)=t(u)I_i(u)\,,\quad \hat{H}(A,B;C,q;u)I_i(u)=0
\ee
where $t(u)\propto\theta(\lambda u)/\theta(u)$ for a given constant $\lambda$ \footnote{In the parametrization given in section \ref{secdeg} for the $S_b^3$ theory, we have $\lambda =e^{2\pi b m_1^{3d}}$.}, 
and the $q$-hypergeometric operator can be written as
\be
\hat{H}(A,B;C,q;u)=\frac{\phantom{}}{\phantom{}}\hat H_2(u)T_q^2+\hat H_1(u)T_q+\hat H_{0}(u);\quad T_qf(u)=f(qu),
\ee
\be
\hat H_2(u)=-\frac{AB-q^{-1}Cu^{-1}}{(q-1)^2}\,,\quad \hat H_1(u)=\frac{(A+B)-(1+q^{-1}C)u^{-1}}{(q-1)^2}\,,\quad \hat H_{0}(u)=-\frac{1-u^{-1}}{(q-1)^2}\,.
\ee
We will now map the $q$-hypergeometric operator to an operator of the form
\be
\hat{A}(u,T_q)=\hat A_1(u)T_q+\hat A_{-1}(u)T_q^{-1}-\hat A_0(u).
\ee
We first compute $\hat A(u,T_q)t(u)$ to get
\be
\hat A(u,T_q)t(u)=t(qu)\hat A_1(u)T_q+t(q^{-1}u)\hat A_{-1}(u)T_q^{-1}-t(u)\hat A_0(u).
\ee
Since $\hat A(u,T_q)=0$ as an operatorial equation, we apply $T_q$ on both sides to get
\ben
T_q\hat A(u,T_q)t(u)&=&t(q^2 u)\left(\hat A_1(qu)T^2_q-\frac{t(qu)}{t(q^2u)}\hat A_0(qu)T_q+\frac{t(u)}{t(q^2u)}\hat A_{-1}(qu)\right).
\een
Imposing
\be
\hat A_1(qu)\sim \hat H_2(u)\,,\quad \lambda \hat A_0(qu)\sim -\hat H_1(u)\,,\quad \lambda^2\hat A_{-1}(qu)\sim H_0(u)\,,
\ee
we are led to identify
\ben
\hat A_1(u)&\sim&-(ABCu^{-1})^{1/2}\left((ABC^{-1})^{1/2}u^{1/2}-(ABC^{-1})^{-1/2}u^{-1/2}\right)\\
\hat A_{-1}(u)&\sim&-\lambda^{-2}(qu^{-1})^{1/2}\left(q^{-1/2}u^{1/2}-q^{1/2}u^{-1/2}\right)\\
\hat A_0(u)&\sim&-\lambda^{-1}(qu^{-1})^{1/2}\left(q^{-1/2}(A+B)u^{1/2}-q^{1/2}(1+q^{-1}C)u^{-1/2}\right).
\een
Upon rescaling, we can finally define
\ben
\hat A_1(u)&=&\frac{(ABC^{-1})^{1/2}u^{1/2}-(ABC^{-1})^{-1/2}u^{-1/2}}{q^{1/2}-q^{-1/2}}\\
\hat A_{-1}(u)&=&\lambda^{-2}(q^{-1}ABC)^{-1/2}\;\frac{q^{-1/2}u^{1/2}-q^{1/2}u^{-1/2}}{q^{1/2}-q^{-1/2}}\\
\hat A_0(u)&=&\lambda^{-1}(q^{-1}ABC)^{-1/2}\;\frac{q^{-1/2}(A+B)u^{1/2}-q^{1/2}(1+q^{-1}C)u^{-1/2}}{q^{1/2}-q^{-1/2}}\,,
\een
so that $\hat A(u,T_q)\mathcal{B}^{\rm 3d}_{1,2}=0$ can be identified with the Baxter equation (see Appendix \ref{BaxtOp}) for the $\mathfrak{sl}(2)$ inhomogeneous XXZ spin-chain of length 1 \cite{XXZref1}, provided 
\be
u=q^v\,,\quad ABC^{-1}=q^{\delta_1+l_1}\,,\quad q^{\delta_1-l_1+1}=1\,,\quad \lambda^2=(q^{-1}ABC)^{-1/2}\,,
\ee
where $v$ is the spectral parameter, while $l_1$ and $\delta_1$ are local parameters on the spin-chain. In this case the blocks $\mathcal{B}^{\rm 3d}_{1,2}$ can be interpreted as eigenvalues of the Baxter $Q$-operator, and $\hat A_0(u)$ as an the eigenvalue of the transfer matrix. 

\bibliography{refs}

\end{document}